\documentclass[11pt]{article}
\usepackage[T1]{fontenc}
\usepackage{jheppub}
\usepackage{wasysym}
\usepackage{color}

\usepackage{tabularx}
\usepackage{physics}
\usepackage{tikz}
\usetikzlibrary{calc}
\usepackage{graphicx} %for images inside equations
\usepackage{ae}
\usetikzlibrary{arrows}
\newcommand{\al}{\tikz \filldraw[scale=0.1, rotate=90] (0,0) -- (1,0) -- (0.5,0.866) -- cycle;}
\newcommand{\ar}{\tikz \filldraw[scale=0.1, rotate=-90] (0,0) -- (1,0) -- (0.5,0.866) -- cycle;}

\usepackage{amsfonts,amsmath,amsthm,amssymb,amsbsy}
\usepackage{mathtools}

\usepackage{pifont}
\title{Wavefunction coefficients from Amplitubes}

\author[]{Ross Glew}\emailAdd{r.glew@herts.ac.uk}
\affiliation[]{Department of Physics, Astronomy and Mathematics, \\ University of Hertfordshire, \\  Hatfield, Hertfordshire, AL10 9AB, United Kingdom}

\abstract{Given a graph its set of connected subgraphs (tubes) can be defined in two ways: either by considering subsets of edges, or by considering subsets of vertices. We refer to these as binary tubes and unary tubes respectively. Both notions come with a natural compatibility condition between tubes which differ by a simple adjacency constraint. Compatible sets of tubes are refered to as tubings. By considering the set of binary tubes, and summing over all maximal binary-tubings, one is lead to an expression for the flat space wavefunction coefficients relevant for computing cosmological correlators. On the other hand, considering the set of unary tubes, and summing over all maximal unary-tubings, one is lead to expressions recently referred to as amplitubes which resemble the scattering amplitudes of $\tr(\phi^3)$ theory. Due to the similarity between these constructions it is natural to expect a close connection between the wavefunction coefficients and amplitubes. In this paper we study the two definitions of tubing in order to provide a new formula for the flat space wavefunction coefficient for a single graph as a sum over products of amplitubes. We also show how the expressions for the amplitubes can naturally be understood as a sum over orientations of the underlying graph. Combining these observations we are lead to an expression for the wavefunction coefficient given by a sum over terms we refer to as decorated amplitubes which matches a recently conjectured formula resulting from partial fractions. Motivated by our rewriting of the wavefunction coefficient we introduce a new definition of tubing which makes use of both the binary and unary tubes which we refer to as cut tubings. We explain how each cut tubing induces a decorated orientation of the underlying graph satisfying an acyclic condition and demonstrate how the set of all acyclic decorated orientations for a given graph count the number of basis functions appearing in the kinematic flow.
}

\begin{document}

\maketitle
%%%%%%%%%%%%%%%%%%%%%%%%%%%%%%%%%%%%%%%%
%%%%%%%%%%%%%%%%%%%%%%%%%%%%%%%%%%%%%%%%
\pagebreak
%%%%%%%%%%%%%%%%%%%%%%%%%%%%%%%%%%%%%%%%%%%%%%%%%%%%%%%%%%%%%%%%%%%
%%%%%%%%%%%%%%%%%%%%%%%%%%%%%%%%%%%%%%%%%%%%%%%%%%%%%%%%%%%%%%%%%%%
\section{Introduction}
Given a graph $G$ there are two natural ways to specify the set of possible subgraphs: either by providing a subset of its edges, or by providing a subset of its vertices. In the case where the subgraphs specified by an edge or vertex set are connected we refer to them as {\it binary} and {\it unary} tubes respectively. The set of binary and unary tubes each come with natural compatibility conditions. Two $b$-tubes are compatible if one is a subgraph of the other or they do not intersect on any vertices \cite{Arkani-Hamed:2017fdk}. Whereas, for the compatibility of $u$-tubes, we have the additional constraint that two $u$-tubes cannot be adjacent on the graph \cite{carr2006coxeter}. Using this compatibility condition $b/u$-tubings are defined as sets of compatible $b/u$-tubes. 

Remarkably, the combinatorial construction of tubings plays a starring role in the computation of physical quantities of interest to cosmologists and particle physicists. In the case of $b$-tubes, by summing over all maximal $b$-tubings, one is lead to the flat space wavefunction coefficients $\Psi_G$. On the other hand, by summing over all maximal $u$-tubings, one is lead to expressions which resemble the scattering amplitudes of $\tr(\phi^3)$ theory, following the terminology introduced in \cite{Glew:2025otn} we refer to these expressions as {\it amplitubes} and denote them by $A_G$. Explicitly the flat space wavefunction and amplitube associated to a graph $G$ are given by
\begin{align}
&\Psi_G = \mathcal{N}_G \sum_{{\bf b} \in \mathcal{B}^{\text{max}}_{G}} \prod_{b \in {\bf b}}\frac{1}{H_{{\bf b}}}, &&A_G = \sum_{{\bf u} \in \mathcal{U}_{G}^{\text{max}} } \prod_{u \in {\bf u}} \frac{1}{H_{\bf u}},
\label{eq:both_tubes}
\end{align}
where $\mathcal{N}_G$ is a normalisation factor, $H_{{\bf b},{\bf u}}$ are functions of the vertices and edges of the graph associated to each $b/u$-tube of $G$ and the sum is over the respective maximal tubings. These expressions are intimately tied to the canonical form of certain positive geometries \cite{Arkani-Hamed:2017tmz}. In the case of the wavefunction coefficients $\Psi_G$ is related to the canonical form of the cosmological polytopes introduced in \cite{Arkani-Hamed:2017fdk}. The amplitubes however are related to the canonical forms of the graph associahedra introduced in \cite{carr2006coxeter}. Both of these have garnered much attention in recent years, see for instance \cite{Arkani-Hamed:2017fdk,Benincasa:2024leu,De:2023xue,Benincasa:2024lxe,De:2024zic,Arkani-Hamed:2023bsv,Arkani-Hamed:2023kig,Benincasa:2020aoj,Arkani-Hamed:2017mur,Arkani-Hamed:2023lbd,He:2020onr,Arkani-Hamed:2024jbp} and references therein. 

At first glance the formulae in \eqref{eq:both_tubes} appear to take a different form. The wavefunction coefficient contains a normalisation factor $\mathcal{N}_G$ given by the product of all edges in the graph, and each term contains $2 |V_E| +L_G-1$ factors in the denominator\footnote{Where $V_G/E_G$ is the vertex/edge set of the graph and $L_G$ is the number of feynman loops.}. Whereas the amplitube has no such prefactor, and each term contains only $|V_E|$ factors in the denominator. However, as we will show, the wavefunction coefficients and amplitubes are intimately connected. 

The connection comes by considering all possible ways of cutting the edges of the graph. If we specify the edges to be cut as ${\bf e} \subset E_G $ we find the wavefunction coefficient can be expanded as a sum over $2^{|E_G|}$ many terms each given by a product of amplitubes. Explicitly this takes the following form
\begin{align}
\Psi_G = \sum_{{\bf e} \subset E_G} (-1)^{|{\bf e}|}  \prod_{b \in {\bf b_e}}  A_{b}.
\label{eq:main_int}
\end{align}
Here the sum is over all subsets ${\bf e}$ of edges of the graph, ${\bf b_{ e}}$ is the $b$-tubing whose $b$-tubes are given by the connected components of $G$ after having cut the edges ${\bf e}$, and $A_{b}$ is the amplitube associated to each connected component. This expansion of the wavefunction coefficient resembles similar formulae arising from the {\it cosmological cutting rules} studied in \cite{Melville:2021lst,AguiSalcedo:2023nds}.

Furthermore, we show how each amplitube can naturally be decomposed into terms corresponding to orientations of the underlying graph. Combining this observation with \eqref{eq:main_int} leads to the following expansion of the wavefunction coefficients
\begin{align}
\Psi_G = \sum_{{\bf e} \subset E_G} (-1)^{|{\bf e}|}  \prod_{b \in {\bf b_e}} \sum_{b^\circ \in b^{\text{dir}}} A_{b^{\circ}}.
\label{eq:main_or}
\end{align}
Here the additional sum is over all valid acyclic orientations of the graph and the factor $A_{b^\circ}$ is the contribution to the amplitube $A_{b}$ from the orientation $b^\circ$. The terms appearing on the right hand side of \eqref{eq:main_or}, which we refer to as {\it decorated amplitubes}, can be labelled by decorating each edge of the graph with one of three options either: a {\it broken} edge depicted as $\begin{tikzpicture}[scale=1]
        \coordinate (A) at (0,0);
        \coordinate (B) at (1/2,0);
        \coordinate (C) at (1,0);
        \coordinate (D) at (3/2,0);
        \draw[thick,dotted] (A) -- (B);
        \fill[black] (A) circle (2pt);
        \fill[black] (B) circle (2pt);
    \end{tikzpicture}$, or with one of the two possible orientations depicited as $\raisebox{-0.11cm}{\begin{tikzpicture}[scale=1]
        \coordinate (A) at (0,0);
        \coordinate (B) at (1/2,0);
        \coordinate (C) at (1,0);
        \coordinate (D) at (3/2,0);
        \draw[thick] (A) --  node {\al} (B);
        \fill[black] (A) circle (2pt);
        \fill[black] (B) circle (2pt);
    \end{tikzpicture}}$ or $\raisebox{-0.11cm}{\begin{tikzpicture}[scale=1]
        \coordinate (A) at (0,0);
        \coordinate (B) at (1/2,0);
        \coordinate (C) at (1,0);
        \coordinate (D) at (3/2,0);
        \draw[thick] (A) --  node {\ar} (B);
        \fill[black] (A) circle (2pt);
        \fill[black] (B) circle (2pt);
    \end{tikzpicture}}$. For instance, for the path graph on three vertices \eqref{eq:main_or} produces a sum over nine terms labelled as 
\begin{align}
\Psi_{\begin{tikzpicture}[scale=0.8]
        \coordinate (A) at (0,0);
        \coordinate (B) at (1/2,0);
        \coordinate (C) at (1,0);
        \coordinate (D) at (3/2,0);
        \draw[thick] (A) -- (B) -- (C) ;
        \fill[black] (A) circle (2pt);
        \fill[black] (B) circle (2pt);
        \fill[black] (C) circle (2pt);
    \end{tikzpicture}}&=A_{\begin{tikzpicture}[scale=0.8]
        \coordinate (A) at (0,0);
        \coordinate (B) at (1/2,0);
        \coordinate (C) at (1,0);
        \coordinate (D) at (3/2,0);
        \draw[thick] (A) -- (B) -- (C) ;
        \draw (A)-- node {\ar} (B);
        \draw (B)-- node {\ar} (C);
        \fill[black] (A) circle (2pt);
        \fill[black] (B) circle (2pt);
        \fill[black] (C) circle (2pt);
    \end{tikzpicture}}+A_{\begin{tikzpicture}[scale=0.8]
        \coordinate (A) at (0,0);
        \coordinate (B) at (1/2,0);
        \coordinate (C) at (1,0);
        \coordinate (D) at (3/2,0);
        \draw[thick] (A) -- (B) -- (C) ;
        \draw (A)-- node {\ar} (B);
        \draw (B)-- node {\al} (C);
        \fill[black] (A) circle (2pt);
        \fill[black] (B) circle (2pt);
        \fill[black] (C) circle (2pt);
    \end{tikzpicture}}+A_{\begin{tikzpicture}[scale=0.8]
        \coordinate (A) at (0,0);
        \coordinate (B) at (1/2,0);
        \coordinate (C) at (1,0);
        \coordinate (D) at (3/2,0);
        \draw[thick] (A) -- (B) -- (C) ;
        \draw (A)-- node {\al} (B);
        \draw (B)-- node {\ar} (C);
        \fill[black] (A) circle (2pt);
        \fill[black] (B) circle (2pt);
        \fill[black] (C) circle (2pt);
    \end{tikzpicture}} \notag \\
    &+A_{\begin{tikzpicture}[scale=0.8]
        \coordinate (A) at (0,0);
        \coordinate (B) at (1/2,0);
        \coordinate (C) at (1,0);
        \coordinate (D) at (3/2,0);
        \draw[thick] (A) -- (B) -- (C) ;
        \draw (A)-- node {\al} (B);
        \draw (B)-- node {\al} (C);
        \fill[black] (A) circle (2pt);
        \fill[black] (B) circle (2pt);
        \fill[black] (C) circle (2pt);
    \end{tikzpicture}}
    -A_{\begin{tikzpicture}[scale=0.8]
        \coordinate (A) at (0,0);
        \coordinate (B) at (1/2,0);
        \coordinate (C) at (1,0);
        \coordinate (D) at (3/2,0);
        \draw[thick] (A) -- (B) ;
        \draw[thick,dotted] (B) -- (C) ;
        \draw (A)-- node {\ar} (B);
        \fill[black] (A) circle (2pt);
        \fill[black] (B) circle (2pt);
        \fill[black] (C) circle (2pt);
    \end{tikzpicture}}-A_{\begin{tikzpicture}[scale=0.8]
        \coordinate (A) at (0,0);
        \coordinate (B) at (1/2,0);
        \coordinate (C) at (1,0);
        \coordinate (D) at (3/2,0);
        \draw[thick] (A) -- (B) ;
        \draw[thick,dotted] (B) -- (C) ;
        \draw (A)-- node {\al} (B);
        \fill[black] (A) circle (2pt);
        \fill[black] (B) circle (2pt);
        \fill[black] (C) circle (2pt);
    \end{tikzpicture}} \notag \\
    &-A_{\begin{tikzpicture}[scale=0.8]
        \coordinate (A) at (0,0);
        \coordinate (B) at (1/2,0);
        \coordinate (C) at (1,0);
        \coordinate (D) at (3/2,0);
        \draw[thick] (B) -- (C) ;
        \draw[thick,dotted] (A) -- (B) ;
        \draw (B)-- node {\ar} (C);
        \fill[black] (A) circle (2pt);
        \fill[black] (B) circle (2pt);
        \fill[black] (C) circle (2pt);
    \end{tikzpicture}}-A_{\begin{tikzpicture}[scale=0.8]
        \coordinate (A) at (0,0);
        \coordinate (B) at (1/2,0);
        \coordinate (C) at (1,0);
        \coordinate (D) at (3/2,0);
        \draw[thick] (B) -- (C) ;
        \draw[thick,dotted] (A) -- (B) ;
        \draw (B)-- node {\al} (C);
        \fill[black] (A) circle (2pt);
        \fill[black] (B) circle (2pt);
        \fill[black] (C) circle (2pt);
    \end{tikzpicture}}+A_{\begin{tikzpicture}[scale=0.8]
        \coordinate (A) at (0,0);
        \coordinate (B) at (1/2,0);x
        \coordinate (C) at (1,0);
        \coordinate (D) at (3/2,0);
        \draw[thick,dotted] (A) -- (B) ;
        \draw[thick,dotted] (B) -- (C) ;
        \fill[black] (A) circle (2pt);
        \fill[black] (B) circle (2pt);
        \fill[black] (C) circle (2pt);
    \end{tikzpicture}}.
\end{align}
This matches the {\it bulk time integral} representation \cite{Arkani-Hamed:2017fdk} of the wavefunction whose general form was recently conjectured in \cite{Fevola:2024nzj} by considering partial fractions. 

Our result \eqref{eq:main_int} for the wavefunction coefficient suggests a hybrid definition of tubing which makes use of both binary and unary tubes, which we refer to as cut tubings, see also \cite{De:2024zic}. Given a cut tubing we show how to assign a {\it decorated orientation} to the underlying graph, with edges receiving one of the following four decorations: $\begin{tikzpicture}[scale=1]
        \coordinate (A) at (0,0);
        \coordinate (B) at (1/2,0);
        \coordinate (C) at (1,0);
        \coordinate (D) at (3/2,0);
        \draw[thick] (A) -- (B);
        \fill[black] (A) circle (2pt);
        \fill[black] (B) circle (2pt);
    \end{tikzpicture}$, $\begin{tikzpicture}[scale=1]
        \coordinate (A) at (0,0);
        \coordinate (B) at (1/2,0);
        \coordinate (C) at (1,0);
        \coordinate (D) at (3/2,0);
        \draw[thick,dotted] (A) -- (B);
        \fill[black] (A) circle (2pt);
        \fill[black] (B) circle (2pt);
    \end{tikzpicture}$, $\raisebox{-0.11cm}{\begin{tikzpicture}[scale=1]
        \coordinate (A) at (0,0);
        \coordinate (B) at (1/2,0);
        \coordinate (C) at (1,0);
        \coordinate (D) at (3/2,0);
        \draw[thick] (A) --  node {\ar} (B);
        \fill[black] (A) circle (2pt);
        \fill[black] (B) circle (2pt);
    \end{tikzpicture}}$ or $\raisebox{-0.11cm}{\begin{tikzpicture}[scale=1]
        \coordinate (A) at (0,0);
        \coordinate (B) at (1/2,0);
        \coordinate (C) at (1,0);
        \coordinate (D) at (3/2,0);
        \draw[thick] (A) --  node {\al} (B);
        \fill[black] (A) circle (2pt);
        \fill[black] (B) circle (2pt);
    \end{tikzpicture}}$. Remarkably, we find that the set of decorated orientations for a given graph satisfying a certain {\it acyclic} condition, denoted $\text{aDec}(G)$, counts the number of basis functions appearing in the kinematic flow \cite{Arkani-Hamed:2023kig,Arkani-Hamed:2023bsv,Baumann:2024mvm}. This leads us to conjecture the following 
\begin{center}
\boxed{\text{ {\it The functions appearing in the kinematic flow for the graph $G$ are counted by $|\text{aDec}(G)|$. }}}
\end{center}
Our rule for determining labels for the set of functions appearing in the kinematic flow for an arbitrary graph $G$ can be mapped to that of \cite{Baumann:2024mvm}. The kinematic flow and its connection to the combinatorics of graph tubings has also been studied in \cite{Grimm:2024mbw,Grimm:2025zhv,He:2024olr,Fan:2024iek,Hang:2024xas}. 
    
 The remainder of the paper is organised as follows. In section \ref{sec:wf} we introduce the wavefunction coefficients $\Psi_G$ and their connection to binary tubes and tubings. In section \ref{sec:amp} we introduce the amplitubes $A_G$ and their connection to unary tubes and tubings. Here we also specify how the amplitube can be decomposed into a sum over oriented graphs. In section \ref{sec:main} we present our main formula \eqref{eq:main} and detail how the wavefunction coefficient can be expanded as a sum over decorated amplitubes. In section \ref{sec:cut_tub} we introduce the notion of cut tubings and decorated orientations and show how the subset of acyclic decorated orientations count the number of basis functions appearing in the kinematic flow.\\[1em]
{\bf Conventions} \\[0.8em]
Throughout the paper we consider a general connected graph $G$ with edge set $E_G$ and vertex set $V_G$. In our definition of graph we allow for {\it loops}, edges whose end points coincide, and multi-edges. We denote the number of feynman loops of the graph by $L_G$. Given a subset of edges ${\bf e} \subset E_G$ we denote the graph induced on the edges of ${\bf e}$ as $G[{\bf e}]$. Note, when the graph contains multi-edges care must be taken to specify which edge is included in the set ${\bf e}$. Similarly, given a subset of vertices ${\bf v} \subset V_{G}$ we denote the graph induced on the vertices of ${\bf v}$ as $G[{\bf v}]$. By definition if both endpoints of an edge $e \in G$ are contained in the subset ${\bf v}$ then the edge is automatically contained in $G[{\bf v}]$. When discussing tubings it will prove useful to completely specify both the vertex and edge set of the subgraph as $\{ {\bf v}, {\bf e}\}$, we denote the graph induced on this edge and vertex set as $G[\{ {\bf v}, {\bf e} \}]$.

%%%%%%%%%%%%%%%%%%%%%%%%%%%%%%%%%%%%%%%%%%%%%%%%%%%%%%%%%%%%%%%%%%%
%%%%%%%%%%%%%%%%%%%%%%%%%%%%%%%%%%%%%%%%%%%%%%%%%%%%%%%%%%%%%%%%%%%
\section{Wavefunction coefficients}
\label{sec:wf}
The flat space wavefunction coefficients can be computed by purely combinatorial means by considering the compatibility of subgraphs as we now review \cite{Arkani-Hamed:2017fdk}. Given a graph $G$ we refer to the set of all {\it connected} subgraphs $G[\{{\bf v},{\bf e} \}]$ obtained by specifying a subset of vertices and edges as {\it binary}\footnote{Where {\it binary} reflects the fact that for each maximal $b$-tubing each $b$-tube is partitioned into exactly {\it two} subtubes. We borrow this terminology from \cite{Balduf:2023tls}.} tubes or $b$-tubes. We denote the set of all $b$-tubes as $B_G$. In simple examples it is useful to introduce a graphical notation for the $b$-tubes by encircling the set of edges and vertices of the graph $G$ contained in the $b$-tube. We say two $b$-tubes are {\it compatible} if one is a subgraph of the other or they do not intersect on any vertices. A set of $b$-tubes ${\bf b}\subset B_G$ is said to form a $b$-tubing if the tubes $b \in {\bf b}$ are mutually compatible. We denote the set of $b$-tubings by $\mathcal{B}_G$. A $b$-tubing is {\it maximal} if no more compatible $b$-tubes can be added. The set of all maximal $b$-tubings is denoted $\mathcal{B}^{\text{max}}_G$. Each maximal $b$-tubing contains exactly $2|V_G|+L-1$ many $b$-tubes.

In terms of $b$-tubings the contribution to the flat space wavefunction coefficient from the graph $G$ (hereby referred to simply as the wavefunction coefficient) can be written as
\begin{align}
&\Psi_G = \left( \prod_{e\in E_G} 2 y_e \right)   \tilde{\Psi}_{G}, \quad \quad  \tilde{\Psi}_{G} = \sum_{{\bf b} \in \mathcal{B}_G^{\text{max}}} \frac{1}{H_{\bf b}}. 
\end{align}
Where the sum is over all maximal $b$-tubings of the graph and we have introduced the notation 
\begin{align}
H_{{\bf b}} = \prod_{b\in {\bf b}} H_{b}.
\end{align}
The $H_b$ are linear functions of the edges (assigned the variables $y_e$) and vertices (assigned the variables $x_v$) of the graph associated to each $b$-tube defined by 
\begin{align}
H_b  = \sum_{v \in b} x_v + \sum_{e \text{ cuts } b} n_{e,b} y_{e}.
\label{eq:lin_func}
\end{align}
Here the first sum is over all vertices contained in the $b$-tube and the second sum is over all edges which are cut by the $b$-tube. The factor $n_{e,b}$ counts the number of times the edge $e$ is cut by the tube $b$.  

The definition of the wavefunction coefficients are best exhibited through examples. For instance for the path graph on three vertices we have 
\begin{align}
\tilde{\Psi}_{\begin{tikzpicture}[scale=0.8]
        \coordinate (A) at (0,0);
        \coordinate (B) at (1/2,0);
        \coordinate (C) at (1,0);
        \coordinate (D) at (3/2,0);
        \draw[thick] (A) -- (B) -- (C);
        \fill[black] (A) circle (2pt);
        \fill[black] (B) circle (2pt);
        \fill[black] (C) circle (2pt);
    \end{tikzpicture}} &=  \frac{1}{H_{ \begin{tikzpicture}[scale=1]
        \coordinate (A) at (0,0);
        \coordinate (B) at (1/2,0);
        \coordinate (C) at (1,0);
        \coordinate (D) at (3/2,0);
        \draw[thick, black] (0.25,0) ellipse (0.41cm and 0.19cm);
        \draw[thick, black] (0.5,0) ellipse (0.75cm and 0.25cm);
        \draw[thick] (A) -- (B) -- (C);
        \draw[black,thick] (A) circle (3pt);
        \draw[black,thick] (B) circle (3pt);
        \draw[black,thick] (C) circle (3pt);
        \fill[black] (A) circle (2pt);
        \fill[black] (B) circle (2pt);
        \fill[black] (C) circle (2pt);
    \end{tikzpicture} }}+\frac{1}{H_{ \begin{tikzpicture}[scale=1]
        \coordinate (A) at (0,0);
        \coordinate (B) at (1/2,0);
        \coordinate (C) at (1,0);
        \coordinate (D) at (3/2,0);
        \draw[thick, black] (0.75,0) ellipse (0.41cm and 0.19cm);
        \draw[thick, black] (0.5,0) ellipse (0.75cm and 0.25cm);
        \draw[thick] (A) -- (B) -- (C);
        \draw[black,thick] (A) circle (3pt);
        \draw[black,thick] (B) circle (3pt);
        \draw[black,thick] (C) circle (3pt);
        \fill[black] (A) circle (2pt);
        \fill[black] (B) circle (2pt);
        \fill[black] (C) circle (2pt);
    \end{tikzpicture} }},
\label{eq:p3}
\end{align}
where we have the following linear factors
\begin{align}
&H_{ \begin{tikzpicture}[scale=0.8]
        \coordinate (A) at (0,0);
        \coordinate (B) at (1/2,0);
        \coordinate (C) at (1,0);
        \coordinate (D) at (3/2,0);
        \draw[thick] (A) -- (B) -- (C);
        \draw[black,thick] (A) circle (4pt);
        \fill[black] (A) circle (2pt);
        \fill[black] (B) circle (2pt);
        \fill[black] (C) circle (2pt);
    \end{tikzpicture} } =x_1 + y_{12},&&H_{ \begin{tikzpicture}[scale=0.8]
        \coordinate (A) at (0,0);
        \coordinate (B) at (1/2,0);
        \coordinate (C) at (1,0);
        \coordinate (D) at (3/2,0);
        \draw[thick] (A) -- (B) -- (C);
        \draw[black,thick] (B) circle (4pt);
        \fill[black] (A) circle (2pt);
        \fill[black] (B) circle (2pt);
        \fill[black] (C) circle (2pt);
    \end{tikzpicture} } =x_2+y_{12}+y_{23},&&&H_{ \begin{tikzpicture}[scale=0.8]
        \coordinate (A) at (0,0);
        \coordinate (B) at (1/2,0);
        \coordinate (C) at (1,0);
        \coordinate (D) at (3/2,0);
        \draw[thick] (A) -- (B) -- (C);
        \draw[black,thick] (C) circle (4pt);
        \fill[black] (A) circle (2pt);
        \fill[black] (B) circle (2pt);
        \fill[black] (C) circle (2pt);
    \end{tikzpicture} } = x_{3}+y_{23}, \notag \\
    &H_{ \begin{tikzpicture}[scale=0.8]
        \coordinate (A) at (0,0);
        \coordinate (B) at (1/2,0);
        \coordinate (C) at (1,0);
        \coordinate (D) at (3/2,0);
        \draw[thick, black] (0.5,0) ellipse (0.75cm and 0.25cm);
        \draw[thick] (A) -- (B) -- (C);
        \fill[black] (A) circle (2pt);
        \fill[black] (B) circle (2pt);
        \fill[black] (C) circle (2pt);
    \end{tikzpicture} } =x_1+x_2+x_3 ,&&H_{ \begin{tikzpicture}[scale=0.8]
        \coordinate (A) at (0,0);
        \coordinate (B) at (1/2,0);
        \coordinate (C) at (1,0);
        \coordinate (D) at (3/2,0);
        \draw[thick, black] (0.25,0) ellipse (0.41cm and 0.19cm);
        \draw[thick] (A) -- (B) -- (C);
        \fill[black] (A) circle (2pt);
        \fill[black] (B) circle (2pt);
        \fill[black] (C) circle (2pt);
    \end{tikzpicture} } =x_1+x_2+y_{23},&&&H_{ \begin{tikzpicture}[scale=0.8]
        \coordinate (A) at (0,0);
        \coordinate (B) at (1/2,0);
        \coordinate (C) at (1,0);
        \coordinate (D) at (3/2,0);
        \draw[thick, black] (0.75,0) ellipse (0.41cm and 0.19cm);
        \draw[thick] (A) -- (B) -- (C);
        \fill[black] (A) circle (2pt);
        \fill[black] (B) circle (2pt);
        \fill[black] (C) circle (2pt);
    \end{tikzpicture} } = x_2+x_3+y_{12}.
\end{align}
The simplest example of a graph containing multi-edges is given by the two-cycle whose flat space wavefunction reads
\begin{align}
\tilde{\Psi}_{
\begin{tikzpicture}[scale=0.6]
\fill[black] (0,-1) circle (2pt);
\fill[black] (1,-1) circle (2pt);
\draw[thick] (0,-1) to[out=90,in=90] (1,-1);
\draw[thick] (0,-1) to[out=-90,in=180+90] (1,-1);
\end{tikzpicture}}
&= \frac{1}{H_{
\begin{tikzpicture}[scale=0.7]
\fill[black] (0,-1) circle (2pt);
\fill[black] (1,-1) circle (2pt);
\draw[black,thick] (1,-1) circle (3.5pt);
\draw[black,thick] (0,-1) circle (3.5pt);
\draw[thick] (0,-1) to[out=90,in=90] (1,-1);
\draw[thick] (0,-1) to[out=-90,in=180+90] (1,-1);
\draw[thick] (0-0.2,-1) to[out=90,in=90] (1+0.2,-1);
\draw[thick] (0+0.2,-1) to[out=90,in=90] (1-0.2,-1);
\draw[thick] (1-0.2,-1) to[out=90+180,in=180] (1,-1.2) to[out=0,in=90+180] (1+0.2,-1);
\draw[thick] (0-0.2,-1) to[out=90+180,in=180] (0,-1.2) to[out=0,in=90+180] (0+0.2,-1);
\draw[thick, black] (0.5,-1) ellipse (0.8cm and 0.5cm);
\end{tikzpicture}}}+ \frac{1}{H_{
\begin{tikzpicture}[scale=0.7]
\fill[black] (0,-1) circle (2pt);
\fill[black] (1,-1) circle (2pt);
\draw[black,thick] (1,-1) circle (3.5pt);
\draw[black,thick] (0,-1) circle (3.5pt);
\draw[thick] (0,-1) to[out=90,in=90] (1,-1);
\draw[thick] (0,-1) to[out=-90,in=180+90] (1,-1);
\draw[thick] (0-0.2,-1) to[out=-90,in=180+90] (1+0.2,-1);
\draw[thick] (0+0.2,-1) to[out=-90,in=180+90] (1-0.2,-1);
\draw[thick] (1-0.2,-1) to[out=90,in=180] (1,-0.8) to[out=0,in=90] (1+0.2,-1);
\draw[thick] (0-0.2,-1) to[out=90,in=180] (0,-0.8) to[out=0,in=90] (0+0.2,-1);
\draw[thick, black] (0.5,-1) ellipse (0.8cm and 0.5cm);
\end{tikzpicture}}},
\label{eq:oneloop}
\end{align}
where we have the following 
\begin{align}
&H_{
\begin{tikzpicture}[scale=0.7]
\fill[black] (0,-1) circle (2pt);
\fill[black] (1,-1) circle (2pt);
\draw[black,thick] (0,-1) circle (4pt);
\draw[thick] (0,-1) to[out=90,in=90] (1,-1);
\draw[thick] (0,-1) to[out=-90,in=180+90] (1,-1);
\end{tikzpicture}}=x_1 + y_{12}+ \tilde{y}_{12}, &&H_{
\begin{tikzpicture}[scale=0.7]
\fill[black] (0,-1) circle (2pt);
\fill[black] (1,-1) circle (2pt);
\draw[black,thick] (1,-1) circle (4pt);
\draw[thick] (0,-1) to[out=90,in=90] (1,-1);
\draw[thick] (0,-1) to[out=-90,in=180+90] (1,-1);
\end{tikzpicture}}=x_2 + y_{12}+ \tilde{y}_{12},&&&H_{
\begin{tikzpicture}[scale=0.7]
\fill[black] (0,-1) circle (2pt);
\fill[black] (1,-1) circle (2pt);
\draw[thick] (0,-1) to[out=90,in=90] (1,-1);
\draw[thick] (0,-1) to[out=-90,in=180+90] (1,-1);
\draw[thick, black] (0.5,-1) ellipse (0.8cm and 0.5cm);
\end{tikzpicture}} = x_1+x_2, \notag \\
&H_{
\begin{tikzpicture}[scale=0.7]
\fill[black] (0,-1) circle (2pt);
\fill[black] (1,-1) circle (2pt);
\draw[thick] (0,-1) to[out=90,in=90] (1,-1);
\draw[thick] (0,-1) to[out=-90,in=180+90] (1,-1);
\draw[thick] (0-0.2,-1) to[out=90,in=90] (1+0.2,-1);
\draw[thick] (0+0.2,-1) to[out=90,in=90] (1-0.2,-1);
\draw[thick] (1-0.2,-1) to[out=90+180,in=180] (1,-1.2) to[out=0,in=90+180] (1+0.2,-1);
\draw[thick] (0-0.2,-1) to[out=90+180,in=180] (0,-1.2) to[out=0,in=90+180] (0+0.2,-1);
\end{tikzpicture}}=x_1+x_2 + 2 \tilde{y}_{12}, &&H_{
\begin{tikzpicture}[scale=0.7]
\fill[black] (0,-1) circle (2pt);
\fill[black] (1,-1) circle (2pt);
\draw[thick] (0,-1) to[out=90,in=90] (1,-1);
\draw[thick] (0,-1) to[out=-90,in=180+90] (1,-1);
\draw[thick] (0-0.2,-1) to[out=-90,in=180+90] (1+0.2,-1);
\draw[thick] (0+0.2,-1) to[out=-90,in=180+90] (1-0.2,-1);
\draw[thick] (1-0.2,-1) to[out=90,in=180] (1,-0.8) to[out=0,in=90] (1+0.2,-1);
\draw[thick] (0-0.2,-1) to[out=90,in=180] (0,-0.8) to[out=0,in=90] (0+0.2,-1);
\end{tikzpicture}}=x_1+x_2 + 2 y_{12}. &&&
\end{align}
To illustrate the final subtlety of graphs containing loops we consider the following example
\begin{align}
\tilde{\Psi}_{\begin{tikzpicture}[scale=1]
        \coordinate (A) at (0,0);
        \coordinate (B) at (1/2,0);
        \draw[thick] (A) -- (B);
        \fill[black] (A) circle (2pt);
        \fill[black] (B) circle (2pt);
        \draw[thick] (B) to[out=90,in=90] (1,0) to[out=90+180,in=90+180] (B);
    \end{tikzpicture}}  = \frac{1}{H_{\begin{tikzpicture}[scale=1]
        \coordinate (A) at (0,0);
        \coordinate (B) at (1/2,0);
        \draw[thick] (A) -- (B);
        \fill[black] (A) circle (2pt);
        \fill[black] (B) circle (2pt);
        \draw[thick] (B) to[out=90,in=90] (1,0) to[out=90+180,in=90+180] (B);
        \draw[thick, black] (0.5,0) ellipse (0.8cm and 0.3cm);
        \draw[thick, black] (0.25,0) ellipse (0.5cm and 0.22cm);
        \draw[black,thick] (A) circle (4pt);
         \draw[black,thick](B) circle (4pt);
    \end{tikzpicture}} }+\frac{1}{H_{\begin{tikzpicture}[scale=1]
        \coordinate (A) at (0,0);
        \coordinate (B) at (1/2,0);
        \draw[thick] (A) -- (B);
        \fill[black] (A) circle (2pt);
        \fill[black] (B) circle (2pt);
        \draw[thick] (B) to[out=90,in=90] (1,0) to[out=90+180,in=90+180] (B);
        \draw[thick, black] (0.5,0) ellipse (0.8cm and 0.3cm);
        \draw[thick, black] (0.7,0) ellipse (0.41cm and 0.23cm);
        \draw[black,thick] (A) circle (4pt);
         \draw[black,thick](B) circle (4pt);
    \end{tikzpicture}} },
\end{align}
where we have the following 
\begin{align}
&H_{\begin{tikzpicture}[scale=1]
        \coordinate (A) at (0,0);
        \coordinate (B) at (1/2,0);
        \draw[thick] (A) -- (B);
        \fill[black] (A) circle (2pt);
        \fill[black] (B) circle (2pt);
        \draw[thick] (B) to[out=90,in=90] (1,0) to[out=90+180,in=90+180] (B);
        \draw[black,thick] (A) circle (4pt);
    \end{tikzpicture}} = x_{1}+y_{12} , &&H_{\begin{tikzpicture}[scale=1]
        \coordinate (A) at (0,0);
        \coordinate (B) at (1/2,0);
        \draw[thick] (A) -- (B);
        \fill[black] (A) circle (2pt);
        \fill[black] (B) circle (2pt);
        \draw[thick] (B) to[out=90,in=90] (1,0) to[out=90+180,in=90+180] (B);
         \draw[black,thick](B) circle (4pt);
    \end{tikzpicture}}=x_2+y_{12}+2y_{22}, &&&H_{\begin{tikzpicture}[scale=1]
        \coordinate (A) at (0,0);
        \coordinate (B) at (1/2,0);
        \draw[thick] (A) -- (B);
        \fill[black] (A) circle (2pt);
        \fill[black] (B) circle (2pt);
        \draw[thick] (B) to[out=90,in=90] (1,0) to[out=90+180,in=90+180] (B);
        \draw[thick, black] (0.5,0) ellipse (0.7cm and 0.3cm);
    \end{tikzpicture}}=x_1+x_2  \notag \\
    &H_{\begin{tikzpicture}[scale=1]
        \coordinate (A) at (0,0);
        \coordinate (B) at (1/2,0);
        \draw[thick] (A) -- (B);
        \fill[black] (A) circle (2pt);
        \fill[black] (B) circle (2pt);
        \draw[thick] (B) to[out=90,in=90] (1,0) to[out=90+180,in=90+180] (B);
        \draw[thick, black] (0.25,0) ellipse (0.5cm and 0.22cm);
    \end{tikzpicture}}=x_{1}+x_2+2y_{22} , &&H_{\begin{tikzpicture}[scale=1]
        \coordinate (A) at (0,0);
        \coordinate (B) at (1/2,0);
        \draw[thick] (A) -- (B);
        \fill[black] (A) circle (2pt);
        \fill[black] (B) circle (2pt);
        \draw[thick] (B) to[out=90,in=90] (1,0) to[out=90+180,in=90+180] (B);
        \draw[thick, black] (0.7,0) ellipse (0.41cm and 0.23cm);
    \end{tikzpicture}}=x_2+y_{12}. &&&
\end{align}
As was mentioned in the introduction the wavefuncion coefficients compute the canonical form of the cosmological polytopes introduced in \cite{Arkani-Hamed:2017fdk}.
%===================================================================
\section{Amplitubes}
\label{sec:amp}
Let us now describe the vertex centric definition of tubes introduced in \cite{carr2006coxeter}. Note, in the case of graphs with multi-edges or loops our definitions differ to those introduced in \cite{carr2011pseudograph}. Let $G$ be a connected graph. A {\it unary}\footnote{Where {\it unary} reflects the fact that for each $u$-tube in a maximal $u$-tubing there is a {\it single} vertex not contained in any of its subtubes.} tube, or $u$-tube, on $G$ is a non-empty subset of vertices of $G$ whose induced subgraph $G[u]$ is connected. In this context we refer to the $u$-tube consisting of the entire vertex set of $G$ as the {\it root}. We will go back and forth between thinking of a $u$-tube as a subgraph or as a subset of vertices. We denote the set of $u$-tubes of $G$ by $U_G$. In the case where the graph is simple we have $U_G = B_G$, generally however, we have $U_G \subset B_G$. Continuing the examples of the last section we have for the path graph 
\begin{align}
U_{\begin{tikzpicture}[scale=0.8]
        \coordinate (A) at (0,0);
        \coordinate (B) at (1/2,0);
        \coordinate (C) at (1,0);
        \draw[thick] (A) -- (B) -- (C);
        \fill[black] (A) circle (2pt);
        \fill[black] (B) circle (2pt);
        \fill[black] (C) circle (2pt);
    \end{tikzpicture}}&=\{ \raisebox{-0.04cm}{\begin{tikzpicture}[scale=1]
        \coordinate (A) at (0,0);
        \coordinate (B) at (1/2,0);
        \coordinate (C) at (1,0);
        \draw[thick] (A) -- (B) -- (C);
        \draw[black,thick] (A) circle (4pt);
        \fill[black] (A) circle (2pt);
        \fill[black] (B) circle (2pt);
        \fill[black] (C) circle (2pt);
    \end{tikzpicture}},\raisebox{-0.04cm}{\begin{tikzpicture}[scale=1]
        \coordinate (A) at (0,0);
        \coordinate (B) at (1/2,0);
        \coordinate (C) at (1,0);
        \draw[thick] (A) -- (B) -- (C);
        \draw[black,thick] (B) circle (4pt);
        \fill[black] (A) circle (2pt);
        \fill[black] (B) circle (2pt);
        \fill[black] (C) circle (2pt);
    \end{tikzpicture}},\raisebox{-0.04cm}{\begin{tikzpicture}[scale=1]
        \coordinate (A) at (0,0);
        \coordinate (B) at (1/2,0);
        \coordinate (C) at (1,0);
        \draw[thick] (A) -- (B) -- (C);
        \draw[black,thick] (C) circle (4pt);
        \fill[black] (A) circle (2pt);
        \fill[black] (B) circle (2pt);
        \fill[black] (C) circle (2pt);
    \end{tikzpicture}} ,\raisebox{-0.09cm}{\begin{tikzpicture}[scale=1]
        \coordinate (A) at (0,0);
        \coordinate (B) at (1/2,0);
        \coordinate (C) at (1,0);
        \draw[thick, black] (0.25,0) ellipse (0.41cm and 0.19cm);
        \draw[thick] (A) -- (B) -- (C);
        \fill[black] (A) circle (2pt);
        \fill[black] (B) circle (2pt);
        \fill[black] (C) circle (2pt);
    \end{tikzpicture}},\raisebox{-0.09cm}{\begin{tikzpicture}[scale=1]
        \coordinate (A) at (0,0);
        \coordinate (B) at (1/2,0);
        \coordinate (C) at (1,0);
        \draw[thick, black] (0.75,0) ellipse (0.41cm and 0.19cm);
        \draw[thick] (A) -- (B) -- (C);
        \fill[black] (A) circle (2pt);
        \fill[black] (B) circle (2pt);
        \fill[black] (C) circle (2pt);
    \end{tikzpicture}},\raisebox{-0.15cm}{\begin{tikzpicture}[scale=1]
        \coordinate (A) at (0,0);
        \coordinate (B) at (1/2,0);
        \coordinate (C) at (1,0);
        \draw[thick, black] (0.5,0) ellipse (0.75cm and 0.25cm);
        \draw[thick] (A) -- (B) -- (C);
        \fill[black] (A) circle (2pt);
        \fill[black] (B) circle (2pt);
        \fill[black] (C) circle (2pt);
    \end{tikzpicture}} \}= B_{\begin{tikzpicture}[scale=0.8]
        \coordinate (A) at (0,0);
        \coordinate (B) at (1/2,0);
        \coordinate (C) at (1,0);
        \draw[thick] (A) -- (B) -- (C);
        \fill[black] (A) circle (2pt);
        \fill[black] (B) circle (2pt);
        \fill[black] (C) circle (2pt);
    \end{tikzpicture}}.
\end{align}
Whilst, for the two-cycle the set of $u$-tubes are given by
\begin{align}
U_{
\begin{tikzpicture}[scale=0.6]
\fill[black] (0,-1) circle (2pt);
\fill[black] (1,-1) circle (2pt);
\draw[thick] (0,-1) to[out=90,in=90] (1,-1);
\draw[thick] (0,-1) to[out=-90,in=180+90] (1,-1);
\end{tikzpicture}} &= \{ 
\raisebox{-0.2cm}{\begin{tikzpicture}[scale=0.8]
\fill[black] (0,-1) circle (2pt);
\fill[black] (1,-1) circle (2pt);
\draw[black,thick] (0,-1) circle (4pt);
\draw[thick] (0,-1) to[out=90,in=90] (1,-1);
\draw[thick] (0,-1) to[out=-90,in=180+90] (1,-1);
\end{tikzpicture}},\raisebox{-0.2cm}{\begin{tikzpicture}[scale=0.8]
\fill[black] (0,-1) circle (2pt);
\fill[black] (1,-1) circle (2pt);
\draw[black,thick] (1,-1) circle (3.5pt);
\draw[thick] (0,-1) to[out=90,in=90] (1,-1);
\draw[thick] (0,-1) to[out=-90,in=180+90] (1,-1);
\end{tikzpicture}},\raisebox{-0.29cm}{\begin{tikzpicture}[scale=0.8]
\fill[black] (0,-1) circle (2pt);
\fill[black] (1,-1) circle (2pt);
\draw[thick] (0,-1) to[out=90,in=90] (1,-1);
\draw[thick] (0,-1) to[out=-90,in=180+90] (1,-1);
\draw[thick, black] (0.5,-1) ellipse (0.8cm and 0.5cm);
\end{tikzpicture}}  \} \subset B_{
\begin{tikzpicture}[scale=0.6]
\fill[black] (0,-1) circle (2pt);
\fill[black] (1,-1) circle (2pt);
\draw[thick] (0,-1) to[out=90,in=90] (1,-1);
\draw[thick] (0,-1) to[out=-90,in=180+90] (1,-1);
\end{tikzpicture}} .
\end{align}
Finally, for the graph containing a loop we have
\begin{align}
U_{\begin{tikzpicture}[scale=1]
        \coordinate (A) at (0,0);
        \coordinate (B) at (1/2,0);
        \draw[thick] (A) -- (B);
        \fill[black] (A) circle (2pt);
        \fill[black] (B) circle (2pt);
        \draw[thick] (B) to[out=90,in=90] (1,0) to[out=90+180,in=90+180] (B);
    \end{tikzpicture}} = \{\raisebox{-0.12cm}{\begin{tikzpicture}[scale=1]
        \coordinate (A) at (0,0);
        \coordinate (B) at (1/2,0);
        \draw[thick] (A) -- (B);
        \fill[black] (A) circle (2pt);
        \fill[black] (B) circle (2pt);
        \draw[thick] (B) to[out=90,in=90] (1,0) to[out=90+180,in=90+180] (B);
        \draw[black,thick] (A) circle (4pt);
    \end{tikzpicture}},\raisebox{-0.15cm}{\begin{tikzpicture}[scale=1]
        \coordinate (A) at (0,0);
        \coordinate (B) at (1/2,0);
        \draw[thick] (A) -- (B);
        \fill[black] (A) circle (2pt);
        \fill[black] (B) circle (2pt);
        \draw[thick] (B) to[out=90,in=90] (1,0) to[out=90+180,in=90+180] (B);
        \draw[thick, black] (0.7,0) ellipse (0.41cm and 0.23cm);
    \end{tikzpicture} },\raisebox{-0.22cm}{\begin{tikzpicture}[scale=1]
        \coordinate (A) at (0,0);
        \coordinate (B) at (1/2,0);
        \draw[thick] (A) -- (B);
        \fill[black] (A) circle (2pt);
        \fill[black] (B) circle (2pt);
        \draw[thick] (B) to[out=90,in=90] (1,0) to[out=90+180,in=90+180] (B);
        \draw[thick, black] (0.5,0) ellipse (0.8cm and 0.3cm);
    \end{tikzpicture}}  \} \subset B_{\begin{tikzpicture}[scale=1]
        \coordinate (A) at (0,0);
        \coordinate (B) at (1/2,0);
        \draw[thick] (A) -- (B);
        \fill[black] (A) circle (2pt);
        \fill[black] (B) circle (2pt);
        \draw[thick] (B) to[out=90,in=90] (1,0) to[out=90+180,in=90+180] (B);
    \end{tikzpicture}} .
\end{align}
We say that two tubes $u_1$ and $u_2$: 
\begin{itemize}
\item {\it intersect} if $u_1 \cap u_2 \neq \emptyset$ and $u_1 \not\subset u_2$ and $u_2 \not\subset u_1$,
\item are {\it adjacent} if $u_1 \cap u_2 = \emptyset $ and $ u_1 \cup u_2 \in U_G$,
\item are {\it compatible} if they do not intersect and they are not adjacent.
\end{itemize}
Compared to the $b$-tubes of the last section we have an additional constraint, the non-adjacency condition, for the compatibility of $u$-tubes. A $u$-tubing ${\bf u} \subset U_G$ is a subset of $u$-tubes which contains the root and whose elements are mutually compatible. As before a {\it maximal} $u$-tubing is a $u$-tubing to which no more compatible $u$-tubes can be added. We denote the set of all $u$-tubings of $G$ by $\mathcal{U}_G$ and the set of all maximal $u$-tubings by $\mathcal{U}^{\text{max}}_G$.  Each maximal $u$-tubing contains exactly $|V_G|$ many $u$-tubes.

Having defined the appropriate graph notions we can proceed by defining the {\it amplitube} \cite{Glew:2025otn} of a graph $G$ as
 \begin{align}
 A_G = \sum_{{\bf u} \in \mathcal{U}_G^{\text{max}}} \frac{1}{H_{{\bf u}}},
 \label{eq:amplitube}
 \end{align} 
where the factors $H_{{\bf u}}$ are the same as those introduced in \eqref{eq:lin_func}. Again the definitions are best illustrated by examples: for the path graph the amplitube is given by
\begin{align}
A_{\begin{tikzpicture}[scale=0.8]
        \coordinate (A) at (0,0);
        \coordinate (B) at (1/2,0);
        \coordinate (C) at (1,0);
        \coordinate (D) at (3/2,0);
        \draw[thick] (A) -- (B) -- (C);
        \fill[black] (A) circle (2pt);
        \fill[black] (B) circle (2pt);
        \fill[black] (C) circle (2pt);
    \end{tikzpicture}} &=  \frac{1}{H_{ \begin{tikzpicture}[scale=1]
        \coordinate (A) at (0,0);
        \coordinate (B) at (1/2,0);
        \coordinate (C) at (1,0);
        \draw[thick, black] (0.25,0) ellipse (0.41cm and 0.19cm);
        \draw[thick, black] (0.5,0) ellipse (0.75cm and 0.25cm);
        \draw[thick] (A) -- (B) -- (C);
        \draw[black,thick] (A) circle (3pt);
        \fill[black] (A) circle (2pt);
        \fill[black] (B) circle (2pt);
        \fill[black] (C) circle (2pt);
    \end{tikzpicture} }}+\frac{1}{H_{ \begin{tikzpicture}[scale=1]
        \coordinate (A) at (0,0);
        \coordinate (B) at (1/2,0);
        \coordinate (C) at (1,0);
        \draw[thick, black] (0.25,0) ellipse (0.41cm and 0.19cm);
        \draw[thick, black] (0.5,0) ellipse (0.75cm and 0.25cm);
        \draw[thick] (A) -- (B) -- (C);
        \draw[black,thick] (B) circle (3pt);
        \fill[black] (A) circle (2pt);
        \fill[black] (B) circle (2pt);
        \fill[black] (C) circle (2pt);
    \end{tikzpicture} }}+\frac{1}{H_{ \begin{tikzpicture}[scale=1]
        \coordinate (A) at (0,0);
        \coordinate (B) at (1/2,0);
        \coordinate (C) at (1,0);
        \coordinate (D) at (3/2,0);
        \draw[thick, black] (0.75,0) ellipse (0.41cm and 0.19cm);
        \draw[thick, black] (0.5,0) ellipse (0.75cm and 0.25cm);
        \draw[thick] (A) -- (B) -- (C);
        \draw[black,thick] (B) circle (3pt);
        \fill[black] (A) circle (2pt);
        \fill[black] (B) circle (2pt);
        \fill[black] (C) circle (2pt);
    \end{tikzpicture}}}+\frac{1}{H_{ \begin{tikzpicture}[scale=1]
        \coordinate (A) at (0,0);
        \coordinate (B) at (1/2,0);
        \coordinate (C) at (1,0);
        \coordinate (D) at (3/2,0);
        \draw[thick, black] (0.75,0) ellipse (0.41cm and 0.19cm);
        \draw[thick, black] (0.5,0) ellipse (0.75cm and 0.25cm);
        \draw[thick] (A) -- (B) -- (C);
        \draw[black,thick] (C) circle (3pt);
        \fill[black] (A) circle (2pt);
        \fill[black] (B) circle (2pt);
        \fill[black] (C) circle (2pt);
    \end{tikzpicture}}}+\frac{1}{H_{ \begin{tikzpicture}[scale=1]
        \coordinate (A) at (0,0);
        \coordinate (B) at (1/2,0);
        \coordinate (C) at (1,0);
        \coordinate (D) at (3/2,0);
        \draw[thick, black] (0.5,0) ellipse (0.75cm and 0.25cm);
        \draw[thick] (A) -- (B) -- (C);
        \draw[black,thick] (A) circle (3pt);
        \draw[black,thick] (C) circle (3pt);
        \fill[black] (A) circle (2pt);
        \fill[black] (B) circle (2pt);
        \fill[black] (C) circle (2pt);
    \end{tikzpicture}}}.
\label{eq:p3}
\end{align}
For the remaining examples introduced in the last section we have the following amplitubes
\begin{align}
A_{
\begin{tikzpicture}[scale=0.6]
\fill[black] (0,-1) circle (2pt);
\fill[black] (1,-1) circle (2pt);
\draw[thick] (0,-1) to[out=90,in=90] (1,-1);
\draw[thick] (0,-1) to[out=-90,in=180+90] (1,-1);
\end{tikzpicture}}
&= \frac{1}{H_{
\begin{tikzpicture}[scale=0.7]
\fill[black] (0,-1) circle (2pt);
\fill[black] (1,-1) circle (2pt);
\draw[black,thick] (0,-1) circle (4pt);
\draw[thick] (0,-1) to[out=90,in=90] (1,-1);
\draw[thick] (0,-1) to[out=-90,in=180+90] (1,-1);
\draw[thick, black] (0.5,-1) ellipse (0.8cm and 0.45cm);
\end{tikzpicture}}}+\frac{1}{H_{
\begin{tikzpicture}[scale=0.7]
\fill[black] (0,-1) circle (2pt);
\fill[black] (1,-1) circle (2pt);
\draw[black,thick] (1,-1) circle (4pt);
\draw[thick] (0,-1) to[out=90,in=90] (1,-1);
\draw[thick] (0,-1) to[out=-90,in=180+90] (1,-1);
\draw[thick, black] (0.5,-1) ellipse (0.8cm and 0.45cm);
\end{tikzpicture}}}, \quad \quad \quad A_{\begin{tikzpicture}[scale=0.8]
        \coordinate (A) at (0,0);
        \coordinate (B) at (1/2,0);
        \draw[thick] (A) -- (B);
        \fill[black] (A) circle (2pt);
        \fill[black] (B) circle (2pt);
        \draw[thick] (B) to[out=90,in=90] (1,0) to[out=90+180,in=90+180] (B);
    \end{tikzpicture}} = \frac{1}{H_{\begin{tikzpicture}[scale=0.8]
        \coordinate (A) at (0,0);
        \coordinate (B) at (1/2,0);
        \draw[thick] (A) -- (B);
        \fill[black] (A) circle (2pt);
        \fill[black] (B) circle (2pt);
        \draw[thick] (B) to[out=90,in=90] (1,0) to[out=90+180,in=90+180] (B);
        \draw[thick, black] (0.5,0) ellipse (0.8cm and 0.3cm);
        \draw[black,thick] (A) circle (4pt);
    \end{tikzpicture}} }+\frac{1}{H_{\begin{tikzpicture}[scale=0.8]
        \coordinate (A) at (0,0);
        \coordinate (B) at (1/2,0);
        \draw[thick] (A) -- (B);
        \fill[black] (A) circle (2pt);
        \fill[black] (B) circle (2pt);
        \draw[thick] (B) to[out=90,in=90] (1,0) to[out=90+180,in=90+180] (B);
        \draw[thick, black] (0.5,0) ellipse (0.8cm and 0.3cm);
        \draw[thick, black] (0.7,0) ellipse (0.41cm and 0.23cm);
    \end{tikzpicture}}}.
\end{align}
We note in passing that the expression for the amplitube can be seen as calculating the canonical form of a convex polytope known as the {\it graph associahedron} \cite{carr2006coxeter} via a sum over its vertices. 

\subsection{Oriented graphs}
We now move on to describe how each maximal $u$-tubing can be used to induce an orientation of the underlying graph and show how this provides a natural decomposition of the corresponding amplitube. Given a graph $G$ and a maximal $u$-tubing ${\bf u}$, for each vertex $v\in V_G$, we can identify a unique tube $u_v^{\uparrow}\in {\bf u}$ defined as the minimal (by inclusion) tube $u \in {\bf u}$ such that $v \in u$. The tube $u_v^\uparrow$ provides a partition of the vertices of $G$ given by
 \begin{align}
V_G= \left( V_{G} \setminus u_v^{\uparrow}\right) \ \cup u_v^{\uparrow}.
 \end{align}
With this partition we can introduce an orientation for the edges of $G$ incident at $v$ by using the following rule:
\begin{align}
 \raisebox{-0.1cm}{\begin{tikzpicture}[scale=1]
        \coordinate (A) at (0,0);
        \coordinate (B) at (1,0);
        \draw[thick] (A) -- (B) ;
        \draw[left] node at (A) {$v$};
        \draw[right] node at (1,0.08) {$v'$};
        \fill[black] (A) circle (2pt);
        \fill[black] (B) circle (2pt);
    \end{tikzpicture}}=\begin{cases}
 \raisebox{-0.1cm}{\begin{tikzpicture}[scale=1]
        \coordinate (A) at (0,0);
        \coordinate (B) at (1,0);
        \draw[thick] (A) -- node {\ar} (B) ;
        \draw[left] node at (A) {$v$};
        \draw[right] node at (1,0.08) {$v'$};
        \fill[black] (A) circle (2pt);
        \fill[black] (B) circle (2pt);
    \end{tikzpicture}} \text{ for } v' \in \left( V_{G} \setminus u_v^{\uparrow}\right),  \\
 \raisebox{-0.1cm}{\begin{tikzpicture}[scale=1]
        \coordinate (A) at (0,0);
        \coordinate (B) at (1,0);
        \draw[thick] (A) --  (B) ;
        \draw[left] node at (A) {$v$};
        \draw[right] node at (1,0.08) {$v'$};
        \fill[black] (A) circle (2pt);
        \fill[black] (B) circle (2pt);
    \end{tikzpicture}} \text{ otherwise }.
\end{cases}
\label{eq:or_rule}
\end{align}
Therefore, for each ${\bf u} \in \mathcal{U}^{\text{max}}_G$, by performing the above procedure for all vertices, we arrive at an orientation for the entire graph which we denote as $G_{{\bf u}}^\circ$. By convention we leave all loops un-oriented. It is straightforward to see that $G_{{\bf u}}^\circ$ is always an acyclic orientation of the graph. We denote the set of all {\it distinct} orientations resulting from this procedure as
\begin{align}
G^{\text{dir}} = \{ G_{{\bf u}}^\circ : {\bf u} \in \mathcal{U}_{G}^{\text{max}}\}.
\end{align}
Since multiple maximal $u$-tubings generally result in the same orientation, it is useful to collect terms in the amplitube as follows
\begin{align}
A_G = \sum_{g^\circ \in G^{\text{dir}}} A_{g^\circ}, \quad \quad \quad  A_{g^\circ} = \sum_{{\bf u}:G^\circ_{{\bf u}}=g^\circ} \frac{1}{H_{\bf u}}.
\label{eq:amp_or}
\end{align}
Where the sum appearing in the second equation is over all maximal tubings ${\bf u}\in \mathcal{U}_{G}^{\text{max}}$ such that $G_{{\bf u}}^{\circ}=g^\circ$. Continuing with our running examples, for the path graph on three vertices \eqref{eq:amp_or} reads
\begin{align}
A_{\begin{tikzpicture}[scale=0.8]
        \coordinate (A) at (0,0);
        \coordinate (B) at (1/2,0);
        \coordinate (C) at (1,0);
        \coordinate (D) at (3/2,0);
        \draw[thick] (A) -- (B) -- (C);
        \fill[black] (A) circle (2pt);
        \fill[black] (B) circle (2pt);
        \fill[black] (C) circle (2pt);
    \end{tikzpicture}} &=  \underbrace{\frac{1}{H_{ \begin{tikzpicture}[scale=1]
        \coordinate (A) at (0,0);
        \coordinate (B) at (1/2,0);
        \coordinate (C) at (1,0);
        \draw[thick, black] (0.25,0) ellipse (0.41cm and 0.19cm);
        \draw[thick, black] (0.5,0) ellipse (0.75cm and 0.25cm);
        \draw[thick] (A) -- (B) -- (C);
        \draw[black,thick] (A) circle (3pt);
        \fill[black] (A) circle (2pt);
        \fill[black] (B) circle (2pt);
        \fill[black] (C) circle (2pt);
    \end{tikzpicture} }}}_{A_{\begin{tikzpicture}[scale=0.8]
        \coordinate (A) at (0,0);
        \coordinate (B) at (1/2,0);
        \coordinate (C) at (1,0);
        \draw[thick] (A) -- (B) -- (C);
        \fill[black] (A) circle (2pt);
        \fill[black] (B) circle (2pt);
        \fill[black] (C) circle (2pt);
        \draw (A)-- node {\ar} (B);
        \draw (B)-- node {\ar} (C);,
    \end{tikzpicture}}}+\underbrace{\frac{1}{H_{ \begin{tikzpicture}[scale=1]
        \coordinate (A) at (0,0);
        \coordinate (B) at (1/2,0);
        \coordinate (C) at (1,0);
        \draw[thick, black] (0.25,0) ellipse (0.41cm and 0.19cm);
        \draw[thick, black] (0.5,0) ellipse (0.75cm and 0.25cm);
        \draw[thick] (A) -- (B) -- (C);
        \draw[black,thick] (B) circle (3pt);
        \fill[black] (A) circle (2pt);
        \fill[black] (B) circle (2pt);
        \fill[black] (C) circle (2pt);
    \end{tikzpicture} }}+\frac{1}{H_{ \begin{tikzpicture}[scale=1]
        \coordinate (A) at (0,0);
        \coordinate (B) at (1/2,0);
        \coordinate (C) at (1,0);
        \coordinate (D) at (3/2,0);
        \draw[thick, black] (0.75,0) ellipse (0.41cm and 0.19cm);
        \draw[thick, black] (0.5,0) ellipse (0.75cm and 0.25cm);
        \draw[thick] (A) -- (B) -- (C);
        \draw[black,thick] (B) circle (3pt);
        \fill[black] (A) circle (2pt);
        \fill[black] (B) circle (2pt);
        \fill[black] (C) circle (2pt);
    \end{tikzpicture}}}}_{A_{\begin{tikzpicture}[scale=1]
        \coordinate (A) at (0,0);
        \coordinate (B) at (1/2,0);
        \coordinate (C) at (1,0);
        \draw[thick] (A) -- (B) -- (C);
        \fill[black] (A) circle (2pt);
        \fill[black] (B) circle (2pt);
        \fill[black] (C) circle (2pt);
        \draw (A)-- node {\al} (B);
        \draw (B)-- node {\ar} (C);,
    \end{tikzpicture}}}+\underbrace{\frac{1}{H_{ \begin{tikzpicture}[scale=1]
        \coordinate (A) at (0,0);
        \coordinate (B) at (1/2,0);
        \coordinate (C) at (1,0);
        \coordinate (D) at (3/2,0);
        \draw[thick, black] (0.5,0) ellipse (0.75cm and 0.25cm);
        \draw[thick] (A) -- (B) -- (C);
        \draw[black,thick] (A) circle (3pt);
        \draw[black,thick] (C) circle (3pt);
        \fill[black] (A) circle (2pt);
        \fill[black] (B) circle (2pt);
        \fill[black] (C) circle (2pt);
    \end{tikzpicture}}}}_{A_{\begin{tikzpicture}[scale=0.8]
        \coordinate (A) at (0,0);
        \coordinate (B) at (1/2,0);
        \coordinate (C) at (1,0);
        \draw[thick] (A) -- (B) -- (C);
        \fill[black] (A) circle (2pt);
        \fill[black] (B) circle (2pt);
        \fill[black] (C) circle (2pt);
        \draw (A)-- node {\ar} (B);
        \draw (B)-- node {\al} (C);,
    \end{tikzpicture}}}+\underbrace{\frac{1}{H_{ \begin{tikzpicture}[scale=1]
        \coordinate (A) at (0,0);
        \coordinate (B) at (1/2,0);
        \coordinate (C) at (1,0);
        \coordinate (D) at (3/2,0);
        \draw[thick, black] (0.75,0) ellipse (0.41cm and 0.19cm);
        \draw[thick, black] (0.5,0) ellipse (0.75cm and 0.25cm);
        \draw[thick] (A) -- (B) -- (C);
        \draw[black,thick] (C) circle (3pt);
        \fill[black] (A) circle (2pt);
        \fill[black] (B) circle (2pt);
        \fill[black] (C) circle (2pt);
    \end{tikzpicture}}}}_{A_{\begin{tikzpicture}[scale=0.8]
        \coordinate (A) at (0,0);
        \coordinate (B) at (1/2,0);
        \coordinate (C) at (1,0);
        \draw[thick] (A) -- (B) -- (C);
        \fill[black] (A) circle (2pt);
        \fill[black] (B) circle (2pt);
        \fill[black] (C) circle (2pt);
        \draw (A)-- node {\al} (B);
        \draw (B)-- node {\al} (C);,
    \end{tikzpicture}}}.
\label{eq:p3}
\end{align}
For the two-cycle we have
\begin{align}
A_{
\begin{tikzpicture}[scale=0.6]
\fill[black] (0,-1) circle (2pt);
\fill[black] (1,-1) circle (2pt);
\draw[thick] (0,-1) to[out=90,in=90] (1,-1);
\draw[thick] (0,-1) to[out=-90,in=180+90] (1,-1);
\end{tikzpicture}}
&=A_{
\begin{tikzpicture}[scale=0.6]
\fill[black] (0,-1) circle (2pt);
\fill[black] (1,-1) circle (2pt);
\draw[thick] (0,-1) to[out=90,in=90] node{\ar}  (1,-1);
\draw[thick] (0,-1) to[out=-90,in=180+90] node{\ar} (1,-1);
\end{tikzpicture}} +A_{
\begin{tikzpicture}[scale=0.6]
\fill[black] (0,-1) circle (2pt);
\fill[black] (1,-1) circle (2pt);
\draw[thick] (0,-1) to[out=90,in=90] node{\al}  (1,-1);
\draw[thick] (0,-1) to[out=-90,in=180+90] node{\al} (1,-1);
\end{tikzpicture}}= \frac{1}{H_{
\begin{tikzpicture}[scale=0.7]
\fill[black] (0,-1) circle (2pt);
\fill[black] (1,-1) circle (2pt);
\draw[black,thick] (0,-1) circle (4pt);
\draw[thick] (0,-1) to[out=90,in=90] (1,-1);
\draw[thick] (0,-1) to[out=-90,in=180+90] (1,-1);
\draw[thick, black] (0.5,-1) ellipse (0.8cm and 0.45cm);
\end{tikzpicture}}}+\frac{1}{H_{
\begin{tikzpicture}[scale=0.7]
\fill[black] (0,-1) circle (2pt);
\fill[black] (1,-1) circle (2pt);
\draw[black,thick] (1,-1) circle (4pt);
\draw[thick] (0,-1) to[out=90,in=90] (1,-1);
\draw[thick] (0,-1) to[out=-90,in=180+90] (1,-1);
\draw[thick, black] (0.5,-1) ellipse (0.8cm and 0.45cm);
\end{tikzpicture}}}.
\end{align}
Finally, for the graph containing a single loop we have 
\begin{align}
A_{\begin{tikzpicture}[scale=0.8]
        \coordinate (A) at (0,0);
        \coordinate (B) at (1/2,0);
        \draw[thick] (A) -- (B);
        \fill[black] (A) circle (2pt);
        \fill[black] (B) circle (2pt);
        \draw[thick] (B) to[out=90,in=90] (1,0) to[out=90+180,in=90+180] (B);
    \end{tikzpicture}} =A_{\begin{tikzpicture}[scale=0.8]
        \coordinate (A) at (0,0);
        \coordinate (B) at (1/2,0);
        \draw[thick] (A) -- node {\ar} (B);
        \fill[black] (A) circle (2pt);
        \fill[black] (B) circle (2pt);
        \draw[thick] (B) to[out=90,in=90] (1,0) to[out=90+180,in=90+180] (B);
    \end{tikzpicture}} +A_{\begin{tikzpicture}[scale=0.8]
        \coordinate (A) at (0,0);
        \coordinate (B) at (1/2,0);
        \draw[thick] (A) -- node {\al} (B);
        \fill[black] (A) circle (2pt);
        \fill[black] (B) circle (2pt);
        \draw[thick] (B) to[out=90,in=90] (1,0) to[out=90+180,in=90+180] (B);
    \end{tikzpicture}} = \frac{1}{H_{\begin{tikzpicture}[scale=0.8]
        \coordinate (A) at (0,0);
        \coordinate (B) at (1/2,0);
        \draw[thick] (A) -- (B);
        \fill[black] (A) circle (2pt);
        \fill[black] (B) circle (2pt);
        \draw[thick] (B) to[out=90,in=90] (1,0) to[out=90+180,in=90+180] (B);
        \draw[thick, black] (0.5,0) ellipse (0.8cm and 0.3cm);
        \draw[black,thick] (A) circle (4pt);
    \end{tikzpicture}} }+\frac{1}{H_{\begin{tikzpicture}[scale=0.8]
        \coordinate (A) at (0,0);
        \coordinate (B) at (1/2,0);
        \draw[thick] (A) -- (B);
        \fill[black] (A) circle (2pt);
        \fill[black] (B) circle (2pt);
        \draw[thick] (B) to[out=90,in=90] (1,0) to[out=90+180,in=90+180] (B);
        \draw[thick, black] (0.5,0) ellipse (0.8cm and 0.3cm);
        \draw[thick, black] (0.7,0) ellipse (0.41cm and 0.23cm);
    \end{tikzpicture}} }.
\end{align}

%======================================
\section{Wavefunction from amplitubes}
\label{sec:main}
In the proceeding sections we have seen how the two notions of tubing, the edge centric binary tubes, and the vertex centric unary tubes, lead to expressions for the flat space wavefunction coefficients and amplitubes respectively. We now move on to study the connection between these two expressions. As we will describe, the wavefunction coefficient can be expanded as a sum over $2^{|E_G|}$ many terms, each associated to cutting a certain subset of edges ${\bf e} \subset E_G$ of the graph. Each term in the sum will correspond to a product of amplitubes, one for each connected component of $G[E_G \setminus {\bf e}]$. The formula we provide is reminiscent to those obtained by the {\it cosmological tree theorem} in \cite{AguiSalcedo:2023nds,Melville:2021lst}.

To begin we introduce the notion of a {\it partition} tubing. Given a subset of edges ${\bf e} \subset E_G$ we define the partition tubing ${\bf b_e}$ as the $b$-tubing with tubes given by the connected components of $G[E_G \setminus {\bf e}]$. In what follows each $b \in {\bf b_e}$ will now play the role of the root $u$-tube for its corresponding subgraph. It will be useful to introduce a graphical notation for the partition tubings by decorating each edge in the corresponding subset ${\bf e}$ by a {\it broken edge} depicted as $\begin{tikzpicture}[scale=1]
        \coordinate (A) at (0,0);
        \coordinate (B) at (1/2,0);
        \coordinate (C) at (1,0);
        \coordinate (D) at (3/2,0);
        \draw[thick,dotted] (A) -- (B);
        \fill[black] (A) circle (2pt);
        \fill[black] (B) circle (2pt);
    \end{tikzpicture}$. Graphically, the partition tubings for our three examples are given by
\begin{center}
\begin{tabular}{cccc}
$\raisebox{-0.16cm}{\begin{tikzpicture}[scale=1]
        \coordinate (A) at (0,0);
        \coordinate (B) at (1/2,0);
        \coordinate (C) at (1,0);
        \draw[thick, black] (0.5,0) ellipse (0.75cm and 0.25cm);
        \draw[thick] (A) -- (B) -- (C);
        \fill[black] (A) circle (2pt);
        \fill[black] (B) circle (2pt);
        \fill[black] (C) circle (2pt);
    \end{tikzpicture}} \leftrightarrow \raisebox{0.03cm}{\begin{tikzpicture}[scale=1]
        \coordinate (A) at (0,0);
        \coordinate (B) at (1/2,0);
        \coordinate (C) at (1,0);
        \draw[thick] (A) -- (B) -- (C);
        \fill[black] (A) circle (2pt);
        \fill[black] (B) circle (2pt);
        \fill[black] (C) circle (2pt);
    \end{tikzpicture}}$, \hfill & $\raisebox{-0.1cm}{\begin{tikzpicture}[scale=1]
        \coordinate (A) at (0,0);
        \coordinate (B) at (1/2,0);
        \coordinate (C) at (1,0);
        \draw[thick, black] (0.25,0) ellipse (0.41cm and 0.19cm);
        \draw[thick] (A) -- (B) -- (C);
        \draw[black,thick] (C) circle (3pt);
        \fill[black] (A) circle (2pt);
        \fill[black] (B) circle (2pt);
        \fill[black] (C) circle (2pt);
    \end{tikzpicture}} \leftrightarrow \raisebox{0.03cm}{\begin{tikzpicture}[scale=1]
        \coordinate (A) at (0,0);
        \coordinate (B) at (1/2,0);
        \coordinate (C) at (1,0);
        \draw[thick] (A) -- (B);
        \draw[thick,dotted] (B) -- (C);
        \fill[black] (A) circle (2pt);
        \fill[black] (B) circle (2pt);
        \fill[black] (C) circle (2pt);
    \end{tikzpicture}}$, \hfill& $\raisebox{-0.1cm}{\begin{tikzpicture}[scale=1]
        \coordinate (A) at (0,0);
        \coordinate (B) at (1/2,0);
        \coordinate (C) at (1,0);
        \draw[thick, black] (0.75,0) ellipse (0.41cm and 0.19cm);
        \draw[thick] (A) -- (B) -- (C);
        \draw[black,thick] (A) circle (3pt);
        \fill[black] (A) circle (2pt);
        \fill[black] (B) circle (2pt);
        \fill[black] (C) circle (2pt);
    \end{tikzpicture}} \leftrightarrow \raisebox{0.03cm}{\begin{tikzpicture}[scale=1]
        \coordinate (A) at (0,0);
        \coordinate (B) at (1/2,0);
        \coordinate (C) at (1,0);
        \draw[thick,dotted] (A) -- (B);
        \draw[thick] (B) -- (C);
        \fill[black] (A) circle (2pt);
        \fill[black] (B) circle (2pt);
        \fill[black] (C) circle (2pt);
    \end{tikzpicture}}$, \hfill& $\raisebox{-0.02cm}{\begin{tikzpicture}[scale=1]
        \coordinate (A) at (0,0);
        \coordinate (B) at (1/2,0);
        \coordinate (C) at (1,0);
        \draw[thick] (A) -- (B) -- (C);
        \draw[black,thick] (A) circle (3pt);
        \draw[black,thick] (B) circle (3pt);
        \draw[black,thick] (C) circle (3pt);
        \fill[black] (A) circle (2pt);
        \fill[black] (B) circle (2pt);
        \fill[black] (C) circle (2pt);
    \end{tikzpicture}} \leftrightarrow \raisebox{0.03cm}{\begin{tikzpicture}[scale=1]
        \coordinate (A) at (0,0);
        \coordinate (B) at (1/2,0);
        \coordinate (C) at (1,0);
        \draw[thick,dotted] (A) -- (B) -- (C);
        \fill[black] (A) circle (2pt);
        \fill[black] (B) circle (2pt);
        \fill[black] (C) circle (2pt);
    \end{tikzpicture}}$,\hfill  \\[0.8em]
    $\raisebox{-0.29cm}{\begin{tikzpicture}[scale=0.8]
\fill[black] (0,-1) circle (2pt);
\fill[black] (1,-1) circle (2pt);
\draw[thick] (0,-1) to[out=90,in=90] (1,-1);
\draw[thick] (0,-1) to[out=-90,in=180+90] (1,-1);
\draw[thick, black] (0.5,-1) ellipse (0.8cm and 0.5cm);
\end{tikzpicture}} \leftrightarrow \raisebox{-0.2cm}{\begin{tikzpicture}[scale=0.8]
\fill[black] (0,-1) circle (2pt);
\fill[black] (1,-1) circle (2pt);
\draw[thick] (0,-1) to[out=90,in=90] (1,-1);
\draw[thick] (0,-1) to[out=-90,in=180+90] (1,-1);
\end{tikzpicture}}$, \hfill & $\raisebox{-0.2cm}{\begin{tikzpicture}[scale=0.8]
\fill[black] (0,-1) circle (2pt);
\fill[black] (1,-1) circle (2pt);
\draw[thick] (0,-1) to[out=90,in=90] (1,-1);
\draw[thick] (0,-1) to[out=-90,in=180+90] (1,-1);
\draw[thick] (0-0.2,-1) to[out=90,in=90] (1+0.2,-1);
\draw[thick] (0+0.2,-1) to[out=90,in=90] (1-0.2,-1);
\draw[thick] (1-0.2,-1) to[out=90+180,in=180] (1,-1.2) to[out=0,in=90+180] (1+0.2,-1);
\draw[thick] (0-0.2,-1) to[out=90+180,in=180] (0,-1.2) to[out=0,in=90+180] (0+0.2,-1);
\end{tikzpicture}} \leftrightarrow \raisebox{-0.2cm}{\begin{tikzpicture}[scale=0.8]
\fill[black] (0,-1) circle (2pt);
\fill[black] (1,-1) circle (2pt);
\draw[thick] (0,-1) to[out=90,in=90] (1,-1);
\draw[thick,dotted] (0,-1) to[out=-90,in=180+90] (1,-1);
\end{tikzpicture}}$, \hfill & $\raisebox{-0.3cm}{\begin{tikzpicture}[scale=0.8]
\fill[black] (0,-1) circle (2pt);
\fill[black] (1,-1) circle (2pt);
\draw[thick] (0,-1) to[out=90,in=90] (1,-1);
\draw[thick] (0,-1) to[out=-90,in=180+90] (1,-1);
\draw[thick] (0-0.2,-1) to[out=-90,in=180+90] (1+0.2,-1);
\draw[thick] (0+0.2,-1) to[out=-90,in=180+90] (1-0.2,-1);
\draw[thick] (1-0.2,-1) to[out=90,in=180] (1,-0.8) to[out=0,in=90] (1+0.2,-1);
\draw[thick] (0-0.2,-1) to[out=90,in=180] (0,-0.8) to[out=0,in=90] (0+0.2,-1);
\end{tikzpicture}} \leftrightarrow \raisebox{-0.2cm}{\begin{tikzpicture}[scale=0.8]
\fill[black] (0,-1) circle (2pt);
\fill[black] (1,-1) circle (2pt);
\draw[thick,dotted] (0,-1) to[out=90,in=90] (1,-1);
\draw[thick] (0,-1) to[out=-90,in=180+90] (1,-1);
\end{tikzpicture}}$, \hfill & $\raisebox{-0.2cm}{\begin{tikzpicture}[scale=0.8]
\fill[black] (0,-1) circle (2pt);
\fill[black] (1,-1) circle (2pt);
\draw[black,thick] (0,-1) circle (4pt);
\draw[black,thick] (1,-1) circle (4pt);
\draw[thick] (0,-1) to[out=90,in=90] (1,-1);
\draw[thick] (0,-1) to[out=-90,in=180+90] (1,-1);
\end{tikzpicture}} \leftrightarrow \raisebox{-0.2cm}{\begin{tikzpicture}[scale=0.8]
\fill[black] (0,-1) circle (2pt);
\fill[black] (1,-1) circle (2pt);
\draw[thick,dotted] (0,-1) to[out=90,in=90] (1,-1);
\draw[thick,dotted] (0,-1) to[out=-90,in=180+90] (1,-1);
\end{tikzpicture}}$, \hfill  \\[0.8em]
    $\raisebox{-0.2cm}{\begin{tikzpicture}[scale=0.9]
        \coordinate (A) at (0,0);
        \coordinate (B) at (1/2,0);
        \draw[thick] (A) -- (B);
        \fill[black] (A) circle (2pt);
        \fill[black] (B) circle (2pt);
        \draw[thick] (B) to[out=90,in=90] (1,0) to[out=90+180,in=90+180] (B);
        \draw[thick, black] (0.5,0) ellipse (0.7cm and 0.3cm);
    \end{tikzpicture}} \leftrightarrow \raisebox{-0.11cm}{\begin{tikzpicture}[scale=0.9]
        \coordinate (A) at (0,0);
        \coordinate (B) at (1/2,0);
        \draw[thick] (A) -- (B);
        \fill[black] (A) circle (2pt);
        \fill[black] (B) circle (2pt);
        \draw[thick] (B) to[out=90,in=90] (1,0) to[out=90+180,in=90+180] (B);
    \end{tikzpicture}}$, \hfill & $\raisebox{-0.13cm}{\begin{tikzpicture}[scale=1]
        \coordinate (A) at (0,0);
        \coordinate (B) at (1/2,0);
        \draw[thick] (A) -- (B);
        \fill[black] (A) circle (2pt);
        \fill[black] (B) circle (2pt);
        \draw[thick] (B) to[out=90,in=90] (1,0) to[out=90+180,in=90+180] (B);
        \draw[thick, black] (0.25,0) ellipse (0.5cm and 0.22cm);
    \end{tikzpicture}} \leftrightarrow \raisebox{-0.1cm}{\begin{tikzpicture}[scale=1]
        \coordinate (A) at (0,0);
        \coordinate (B) at (1/2,0);
        \draw[thick] (A) -- (B);
        \fill[black] (A) circle (2pt);
        \fill[black] (B) circle (2pt);
        \draw[thick,dotted] (B) to[out=90,in=90] (1,0) to[out=90+180,in=90+180] (B);
    \end{tikzpicture}}$, \hfill & $\raisebox{-0.12cm}{\begin{tikzpicture}[scale=1]
        \coordinate (A) at (0,0);
        \coordinate (B) at (1/2,0);
        \draw[thick] (A) -- (B);
        \fill[black] (A) circle (2pt);
        \fill[black] (B) circle (2pt);
        \draw[black,thick] (A) circle (4pt);
        \draw[thick] (B) to[out=90,in=90] (1,0) to[out=90+180,in=90+180] (B);
        \draw[thick, black] (0.7,0) ellipse (0.41cm and 0.23cm);
    \end{tikzpicture}} \leftrightarrow \raisebox{-0.1cm}{\begin{tikzpicture}[scale=1]
        \coordinate (A) at (0,0);
        \coordinate (B) at (1/2,0);
        \draw[thick,dotted] (A) -- (B);
        \fill[black] (A) circle (2pt);
        \fill[black] (B) circle (2pt);
        \draw[thick] (B) to[out=90,in=90] (1,0) to[out=90+180,in=90+180] (B);
    \end{tikzpicture}}$, \hfill & $\raisebox{-0.1cm}{\begin{tikzpicture}[scale=1]
        \coordinate (A) at (0,0);
        \coordinate (B) at (1/2,0);
        \draw[thick] (A) -- (B);
        \fill[black] (A) circle (2pt);
        \fill[black] (B) circle (2pt);
        \draw[black,thick] (A) circle (4pt);
        \draw[black,thick] (B) circle (4pt);
        \draw[thick] (B) to[out=90,in=90] (1,0) to[out=90+180,in=90+180] (B);
    \end{tikzpicture}} \leftrightarrow \raisebox{-0.1cm}{\begin{tikzpicture}[scale=1]
        \coordinate (A) at (0,0);
        \coordinate (B) at (1/2,0);
        \draw[thick,dotted] (A) -- (B);
        \fill[black] (A) circle (2pt);
        \fill[black] (B) circle (2pt);
        \draw[thick,dotted] (B) to[out=90,in=90] (1,0) to[out=90+180,in=90+180] (B);
    \end{tikzpicture}}$. \hfill
\end{tabular}
\label{eq:tab}
\end{center}
Having introduced this notation we find that the wavefunction coefficient can be expanded as a sum over partition tubings as 
\begin{align}
\Psi_G = \sum_{{\bf e} \subset E_G} (-1)^{|{\bf e}|}  \prod_{b \in {\bf b_e}} A_{b}.
\label{eq:edge_expand}
\end{align}
The contribution from each partition tubing ${\bf b}_e$ is given simply by the product of the amplitubes associated to each subgraph $b \in {\bf b}_e$. Note, in particular each term appearing in the sum now has only $|V_G|$ many factors in the denominator. We have checked explicitly \eqref{eq:edge_expand} in a large number of cases including examples with loops and multi-edges.

As an illustration of formula \eqref{eq:edge_expand} for our running examples we have: for the path graph 
\begin{align}
\Psi_{\begin{tikzpicture}[scale=0.8]
        \coordinate (A) at (0,0);
        \coordinate (B) at (1/2,0);
        \coordinate (C) at (1,0);
        \coordinate (D) at (3/2,0);
        \draw[thick] (A) -- (B) -- (C) ;
        \fill[black] (A) circle (2pt);
        \fill[black] (B) circle (2pt);
        \fill[black] (C) circle (2pt);
    \end{tikzpicture}}=A_{\begin{tikzpicture}[scale=0.8]
        \coordinate (A) at (0,0);
        \coordinate (B) at (1/2,0);
        \coordinate (C) at (1,0);
        \coordinate (D) at (3/2,0);
        \draw[thick] (A) -- (B) -- (C) ;
        \fill[black] (A) circle (2pt);
        \fill[black] (B) circle (2pt);
        \fill[black] (C) circle (2pt);
    \end{tikzpicture}}-A_{\begin{tikzpicture}[scale=0.8]
        \coordinate (A) at (0,0);
        \coordinate (B) at (1/2,0);
        \coordinate (C) at (1,0);
        \coordinate (D) at (3/2,0);
        \draw[thick] (A) -- (B);
        \draw[thick,dotted] (B) -- (C) ;
        \fill[black] (A) circle (2pt);
        \fill[black] (B) circle (2pt);
        \fill[black] (C) circle (2pt);
    \end{tikzpicture}}-A_{\begin{tikzpicture}[scale=0.8]
        \coordinate (A) at (0,0);
        \coordinate (B) at (1/2,0);
        \coordinate (C) at (1,0);
        \coordinate (D) at (3/2,0);
        \draw[thick,dotted] (A) -- (B);
        \draw[thick] (B) -- (C) ;
        \fill[black] (A) circle (2pt);
        \fill[black] (B) circle (2pt);
        \fill[black] (C) circle (2pt);
    \end{tikzpicture}}+A_{\begin{tikzpicture}[scale=0.8]
        \coordinate (A) at (0,0);
        \coordinate (B) at (1/2,0);
        \coordinate (C) at (1,0);
        \coordinate (D) at (3/2,0);
        \draw[thick,dotted] (A) -- (B) -- (C) ;
        \fill[black] (A) circle (2pt);
        \fill[black] (B) circle (2pt);
        \fill[black] (C) circle (2pt);
    \end{tikzpicture}},
    \label{eq:edge_expand_path}
\end{align}
for the two-cycle
\begin{align}
\Psi_{
\begin{tikzpicture}[scale=0.6]
\fill[black] (0,-1) circle (2pt);
\fill[black] (1,-1) circle (2pt);
\draw[thick] (0,-1) to[out=90,in=90] (1,-1);
\draw[thick] (0,-1) to[out=-90,in=180+90] (1,-1);
\end{tikzpicture}}=&A_{
\begin{tikzpicture}[scale=0.6]
\fill[black] (0,-1) circle (2pt);
\fill[black] (1,-1) circle (2pt);
\draw[thick] (0,-1) to[out=90,in=90] (1,-1);
\draw[thick] (0,-1) to[out=-90,in=180+90] (1,-1);
\end{tikzpicture}}-A_{
\begin{tikzpicture}[scale=0.6]
\fill[black] (0,-1) circle (2pt);
\fill[black] (1,-1) circle (2pt);
\draw[thick] (0,-1) to[out=90,in=90] (1,-1);
\draw[thick,dotted] (0,-1) to[out=-90,in=180+90] (1,-1);
\end{tikzpicture}}
-A_{
\begin{tikzpicture}[scale=0.6]
\fill[black] (0,-1) circle (2pt);
\fill[black] (1,-1) circle (2pt);
\draw[thick,dotted] (0,-1) to[out=90,in=90]  (1,-1);
\draw[thick] (0,-1) to[out=-90,in=180+90] (1,-1);
\end{tikzpicture}}+A_{
\begin{tikzpicture}[scale=0.6]
\fill[black] (0,-1) circle (2pt);
\fill[black] (1,-1) circle (2pt);
\draw[thick,dotted] (0,-1) to[out=90,in=90]  (1,-1);
\draw[thick,dotted] (0,-1) to[out=-90,in=180+90] (1,-1);
\end{tikzpicture}},
\label{eq:edge_expand_cyc}
\end{align}
and for our last example
\begin{align}
\Psi_{\begin{tikzpicture}[scale=0.8]
        \coordinate (A) at (0,0);
        \coordinate (B) at (1/2,0);
        \draw[thick] (A) -- (B);
        \fill[black] (A) circle (2pt);
        \fill[black] (B) circle (2pt);
        \draw[thick] (B) to[out=90,in=90] (1,0) to[out=90+180,in=90+180] (B);
    \end{tikzpicture}} =A_{\begin{tikzpicture}[scale=0.8]
        \coordinate (A) at (0,0);
        \coordinate (B) at (1/2,0);
        \draw[thick] (A) -- (B);
        \fill[black] (A) circle (2pt);
        \fill[black] (B) circle (2pt);
        \draw[thick] (B) to[out=90,in=90] (1,0) to[out=90+180,in=90+180] (B);
    \end{tikzpicture}}  -A_{\begin{tikzpicture}[scale=0.8]
        \coordinate (A) at (0,0);
        \coordinate (B) at (1/2,0);
        \draw[thick] (A) --  (B);
        \fill[black] (A) circle (2pt);
        \fill[black] (B) circle (2pt);
        \draw[thick,dotted] (B) to[out=90,in=90] (1,0) to[out=90+180,in=90+180] (B);
    \end{tikzpicture}} -A_{\begin{tikzpicture}[scale=0.8]
        \coordinate (A) at (0,0);
        \coordinate (B) at (1/2,0);
        \draw[thick,dotted] (A) --  (B);
        \fill[black] (A) circle (2pt);
        \fill[black] (B) circle (2pt);
        \draw[thick] (B) to[out=90,in=90] (1,0) to[out=90+180,in=90+180] (B);
    \end{tikzpicture}} +A_{\begin{tikzpicture}[scale=0.8]
        \coordinate (A) at (0,0);
        \coordinate (B) at (1/2,0);
        \draw[thick,dotted] (A) -- (B);
        \fill[black] (A) circle (2pt);
        \fill[black] (B) circle (2pt);
        \draw[thick,dotted] (B) to[out=90,in=90] (1,0) to[out=90+180,in=90+180] (B);
    \end{tikzpicture}}.
    \label{eq:edge_expand_loop}
\end{align}
The terms appearing in \eqref{eq:edge_expand_path} with broken edges are given by the following products of amplitubes
\begin{align}
A_{\begin{tikzpicture}[scale=0.8]
        \coordinate (A) at (0,0);
        \coordinate (B) at (1/2,0);
        \coordinate (C) at (1,0);
        \coordinate (D) at (3/2,0);
        \draw[thick,dotted] (B) -- (C) ;
        \draw[thick] (A) -- (B) ;
        \fill[black] (A) circle (2pt);
        \fill[black] (B) circle (2pt);
        \fill[black] (C) circle (2pt);
    \end{tikzpicture}} &=\left(\frac{1}{H_{ \begin{tikzpicture}[scale=0.8]
        \coordinate (A) at (0,0);
        \coordinate (B) at (1/2,0);
        \coordinate (C) at (1,0);
        \draw[thick, black] (0.25,0) ellipse (0.47cm and 0.25cm);
        \draw[thick] (A) -- (B) -- (C);
        \draw[black,thick] (A) circle (4pt);
        \fill[black] (A) circle (2pt);
        \fill[black] (B) circle (2pt);
        \fill[black] (C) circle (2pt);
    \end{tikzpicture} }}+\frac{1}{H_{ \begin{tikzpicture}[scale=0.8]
        \coordinate (A) at (0,0);
        \coordinate (B) at (1/2,0);
        \coordinate (C) at (1,0);
        \draw[thick, black] (0.25,0) ellipse (0.47cm and 0.25cm);
        \draw[thick] (A) -- (B) -- (C);
        \draw[black,thick] (B) circle (4pt);
        \fill[black] (A) circle (2pt);
        \fill[black] (B) circle (2pt);
        \fill[black] (C) circle (2pt);
    \end{tikzpicture} }} \right) \times \frac{1}{H_{ \begin{tikzpicture}[scale=0.8]
        \coordinate (A) at (0,0);
        \coordinate (B) at (1/2,0);
        \coordinate (C) at (1,0);
        \draw[thick] (A) -- (B) -- (C);
        \draw[black,thick] (C) circle (4pt);
        \fill[black] (A) circle (2pt);
        \fill[black] (B) circle (2pt);
        \fill[black] (C) circle (2pt);
    \end{tikzpicture} }}, \notag \\ 
    A_{\begin{tikzpicture}[scale=0.8]
        \coordinate (A) at (0,0);
        \coordinate (B) at (1/2,0);
        \coordinate (C) at (1,0);
        \coordinate (D) at (3/2,0);
        \draw[thick,dotted] (A) -- (B) ;
        \draw[thick] (B) -- (C) ;
        \fill[black] (A) circle (2pt);
        \fill[black] (B) circle (2pt);
        \fill[black] (C) circle (2pt);
    \end{tikzpicture}}&=\frac{1}{H_{ \begin{tikzpicture}[scale=0.8]
        \coordinate (A) at (0,0);
        \coordinate (B) at (1/2,0);
        \coordinate (C) at (1,0);
        \draw[thick] (A) -- (B) -- (C);
        \draw[black,thick] (A) circle (4pt);
        \fill[black] (A) circle (2pt);
        \fill[black] (B) circle (2pt);
        \fill[black] (C) circle (2pt);
    \end{tikzpicture} }}\times \left(\frac{1}{H_{ \begin{tikzpicture}[scale=0.8]
        \coordinate (A) at (0,0);
        \coordinate (B) at (1/2,0);
        \coordinate (C) at (1,0);
        \draw[thick, black] (0.75,0) ellipse (0.47cm and 0.25cm);
        \draw[thick] (A) -- (B) -- (C);
        \draw[black,thick] (B) circle (4pt);
        \fill[black] (A) circle (2pt);
        \fill[black] (B) circle (2pt);
        \fill[black] (C) circle (2pt);
    \end{tikzpicture} }}+\frac{1}{H_{ \begin{tikzpicture}[scale=0.8]
        \coordinate (A) at (0,0);
        \coordinate (B) at (1/2,0);
        \coordinate (C) at (1,0);
        \draw[thick, black] (0.75,0) ellipse (0.47cm and 0.25cm);
        \draw[thick] (A) -- (B) -- (C);
        \draw[black,thick] (C) circle (4pt);
        \fill[black] (A) circle (2pt);
        \fill[black] (B) circle (2pt);
        \fill[black] (C) circle (2pt);
    \end{tikzpicture} }} \right), \notag \\ 
A_{\begin{tikzpicture}[scale=0.8]
        \coordinate (A) at (0,0);
        \coordinate (B) at (1/2,0);
        \coordinate (C) at (1,0);
        \coordinate (D) at (3/2,0);
        \draw[thick,dotted] (A) -- (B) -- (C) ;
        \fill[black] (A) circle (2pt);
        \fill[black] (B) circle (2pt);
        \fill[black] (C) circle (2pt);
    \end{tikzpicture}}&=\frac{1}{H_{ \begin{tikzpicture}[scale=0.8]
        \coordinate (A) at (0,0);
        \coordinate (B) at (1/2,0);
        \coordinate (C) at (1,0);
        \draw[thick] (A) -- (B) -- (C);
        \draw[black,thick] (A) circle (4pt);
        \fill[black] (A) circle (2pt);
        \fill[black] (B) circle (2pt);
        \fill[black] (C) circle (2pt);
    \end{tikzpicture} }}\times \frac{1}{H_{ \begin{tikzpicture}[scale=0.8]
        \coordinate (A) at (0,0);
        \coordinate (B) at (1/2,0);
        \coordinate (C) at (1,0);
        \draw[thick] (A) -- (B) -- (C);
        \draw[black,thick] (B) circle (4pt);
        \fill[black] (A) circle (2pt);
        \fill[black] (B) circle (2pt);
        \fill[black] (C) circle (2pt);
    \end{tikzpicture} }} \times \frac{1}{H_{ \begin{tikzpicture}[scale=0.8]
        \coordinate (A) at (0,0);
        \coordinate (B) at (1/2,0);
        \coordinate (C) at (1,0);
        \draw[thick] (A) -- (B) -- (C);
        \draw[black,thick] (C) circle (4pt);
        \fill[black] (A) circle (2pt);
        \fill[black] (B) circle (2pt);
        \fill[black] (C) circle (2pt);
    \end{tikzpicture} }}.
\end{align}
For the two-cycle the terms appearing in \eqref{eq:edge_expand_cyc} with broken edges are given by the following products of amplitubes
\begin{align}
A_{
\begin{tikzpicture}[scale=0.55]
\fill[black] (0,-1) circle (2pt);
\fill[black] (1,-1) circle (2pt);
\draw[thick] (0,-1) to[out=90,in=90] (1,-1);
\draw[thick,dotted] (0,-1) to[out=-90,in=180+90] (1,-1);
\end{tikzpicture}} &=\frac{1}{H_{
\begin{tikzpicture}[scale=0.55]
\fill[black] (0,-1) circle (2pt);
\fill[black] (1,-1) circle (2pt);
\draw[black,thick] (0,-1) circle (3.5pt);
\draw[thick] (0,-1) to[out=90,in=90] (1,-1);
\draw[thick] (0,-1) to[out=-90,in=180+90] (1,-1);
\draw[thick] (0-0.2,-1) to[out=90,in=90] (1+0.2,-1);
\draw[thick] (0+0.2,-1) to[out=90,in=90] (1-0.2,-1);
\draw[thick] (1-0.2,-1) to[out=90+180,in=180] (1,-1.2) to[out=0,in=90+180] (1+0.2,-1);
\draw[thick] (0-0.2,-1) to[out=90+180,in=180] (0,-1.2) to[out=0,in=90+180] (0+0.2,-1);
\end{tikzpicture}}}+\frac{1}{H_{
\begin{tikzpicture}[scale=0.55]
\fill[black] (0,-1) circle (2pt);
\fill[black] (1,-1) circle (2pt);
\draw[black,thick] (1,-1) circle (3.5pt);
\draw[thick] (0,-1) to[out=90,in=90] (1,-1);
\draw[thick] (0,-1) to[out=-90,in=180+90] (1,-1);
\draw[thick] (0-0.2,-1) to[out=90,in=90] (1+0.2,-1);
\draw[thick] (0+0.2,-1) to[out=90,in=90] (1-0.2,-1);
\draw[thick] (1-0.2,-1) to[out=90+180,in=180] (1,-1.2) to[out=0,in=90+180] (1+0.2,-1);
\draw[thick] (0-0.2,-1) to[out=90+180,in=180] (0,-1.2) to[out=0,in=90+180] (0+0.2,-1);
\end{tikzpicture}}}, \quad \quad  
A_{
\begin{tikzpicture}[scale=0.55]
\fill[black] (0,-1) circle (2pt);
\fill[black] (1,-1) circle (2pt);
\draw[thick,dotted] (0,-1) to[out=90,in=90]  (1,-1);
\draw[thick] (0,-1) to[out=-90,in=180+90] (1,-1);
\end{tikzpicture}}=\frac{1}{H_{
\begin{tikzpicture}[scale=0.55]
\fill[black] (0,-1) circle (2pt);
\fill[black] (1,-1) circle (2pt);
\draw[black,thick] (0,-1) circle (3.5pt);
\draw[thick] (0,-1) to[out=90,in=90] (1,-1);
\draw[thick] (0,-1) to[out=-90,in=180+90] (1,-1);
\draw[thick] (0-0.2,-1) to[out=-90,in=180+90] (1+0.2,-1);
\draw[thick] (0+0.2,-1) to[out=-90,in=180+90] (1-0.2,-1);
\draw[thick] (1-0.2,-1) to[out=90,in=180] (1,-0.8) to[out=0,in=90] (1+0.2,-1);
\draw[thick] (0-0.2,-1) to[out=90,in=180] (0,-0.8) to[out=0,in=90] (0+0.2,-1);
\end{tikzpicture}}}+\frac{1}{H_{
\begin{tikzpicture}[scale=0.55]
\fill[black] (0,-1) circle (2pt);
\fill[black] (1,-1) circle (2pt);
\draw[black,thick] (1,-1) circle (3.5pt);
\draw[thick] (0,-1) to[out=90,in=90] (1,-1);
\draw[thick] (0,-1) to[out=-90,in=180+90] (1,-1);
\draw[thick] (0-0.2,-1) to[out=-90,in=180+90] (1+0.2,-1);
\draw[thick] (0+0.2,-1) to[out=-90,in=180+90] (1-0.2,-1);
\draw[thick] (1-0.2,-1) to[out=90,in=180] (1,-0.8) to[out=0,in=90] (1+0.2,-1);
\draw[thick] (0-0.2,-1) to[out=90,in=180] (0,-0.8) to[out=0,in=90] (0+0.2,-1);
\end{tikzpicture}}}, \quad \quad 
A_{
\begin{tikzpicture}[scale=0.55]
\fill[black] (0,-1) circle (2pt);
\fill[black] (1,-1) circle (2pt);
\draw[thick,dotted] (0,-1) to[out=90,in=90]  (1,-1);
\draw[thick,dotted] (0,-1) to[out=-90,in=180+90]  (1,-1);
\end{tikzpicture}}=\frac{1}{H_{
\begin{tikzpicture}[scale=0.6]
\fill[black] (0,-1) circle (2pt);
\fill[black] (1,-1) circle (2pt);
\draw[black,thick] (0,-1) circle (4pt);
\draw[thick] (0,-1) to[out=90,in=90] (1,-1);
\draw[thick] (0,-1) to[out=-90,in=180+90] (1,-1);
\end{tikzpicture}}} \times \frac{1}{H_{
\begin{tikzpicture}[scale=0.55]
\fill[black] (0,-1) circle (2pt);
\fill[black] (1,-1) circle (2pt);
\draw[black,thick] (1,-1) circle (4pt);
\draw[thick] (0,-1) to[out=90,in=90] (1,-1);
\draw[thick] (0,-1) to[out=-90,in=180+90] (1,-1);
\end{tikzpicture}}}. \notag
\end{align}
Finally, the terms appearing in \eqref{eq:edge_expand_loop} with broken edges are given by the following products of amplitubes
\begin{align}
A_{\begin{tikzpicture}[scale=0.8]
        \coordinate (A) at (0,0);
        \coordinate (B) at (1/2,0);
        \draw[thick] (A) --  (B);
        \fill[black] (A) circle (2pt);
        \fill[black] (B) circle (2pt);
        \draw[thick,dotted] (B) to[out=90,in=90] (1,0) to[out=90+180,in=90+180] (B);
    \end{tikzpicture}}  &=\frac{1}{H_{\begin{tikzpicture}[scale=0.8]
        \coordinate (A) at (0,0);
        \coordinate (B) at (1/2,0);
        \draw[thick] (A) -- (B);
        \fill[black] (A) circle (2pt);
        \fill[black] (B) circle (2pt);
        \draw[thick] (B) to[out=90,in=90] (1,0) to[out=90+180,in=90+180] (B);
        \draw[thick, black] (0.25,0) ellipse (0.5cm and 0.22cm);
        \draw[black,thick] (A) circle (4pt);
    \end{tikzpicture}} }+\frac{1}{H_{\begin{tikzpicture}[scale=0.8]
        \coordinate (A) at (0,0);
        \coordinate (B) at (1/2,0);
        \draw[thick] (A) -- (B);
        \fill[black] (A) circle (2pt);
        \fill[black] (B) circle (2pt);
        \draw[thick] (B) to[out=90,in=90] (1,0) to[out=90+180,in=90+180] (B);
        \draw[thick, black] (0.25,0) ellipse (0.5cm and 0.22cm);
        \draw[black,thick] (B) circle (4pt);
    \end{tikzpicture}} }, \quad \quad \quad  A_{\begin{tikzpicture}[scale=0.8]
        \coordinate (A) at (0,0);
        \coordinate (B) at (1/2,0);
        \draw[thick,dotted] (A) --  (B);
        \fill[black] (A) circle (2pt);
        \fill[black] (B) circle (2pt);
        \draw[thick] (B) to[out=90,in=90] (1,0) to[out=90+180,in=90+180] (B);
    \end{tikzpicture}}=\frac{1}{H_{\begin{tikzpicture}[scale=0.8]
        \coordinate (A) at (0,0);
        \coordinate (B) at (1/2,0);
        \draw[thick] (A) -- (B);
        \fill[black] (A) circle (2pt);
        \fill[black] (B) circle (2pt);
        \draw[thick] (B) to[out=90,in=90] (1,0) to[out=90+180,in=90+180] (B);
        \draw[black,thick] (A) circle (4pt);
    \end{tikzpicture}}} \times \frac{1}{H_{\begin{tikzpicture}[scale=0.8]
        \coordinate (A) at (0,0);
        \coordinate (B) at (1/2,0);
        \draw[thick] (A) -- (B);
        \fill[black] (A) circle (2pt);
        \fill[black] (B) circle (2pt);
        \draw[thick] (B) to[out=90,in=90] (1,0) to[out=90+180,in=90+180] (B);
        \draw[thick, black] (0.7,0) ellipse (0.41cm and 0.23cm);
    \end{tikzpicture}}}, \notag \\
     A_{\begin{tikzpicture}[scale=0.8]
        \coordinate (A) at (0,0);
        \coordinate (B) at (1/2,0);
        \draw[thick,dotted] (A) -- (B);
        \fill[black] (A) circle (2pt);
        \fill[black] (B) circle (2pt);
        \draw[thick,dotted] (B) to[out=90,in=90] (1,0) to[out=90+180,in=90+180] (B);
    \end{tikzpicture}}&=\frac{1}{H_{\begin{tikzpicture}[scale=0.8]
        \coordinate (A) at (0,0);
        \coordinate (B) at (1/2,0);
        \draw[thick] (A) -- (B);
        \fill[black] (A) circle (2pt);
        \fill[black] (B) circle (2pt);
        \draw[thick] (B) to[out=90,in=90] (1,0) to[out=90+180,in=90+180] (B);
        \draw[black,thick] (A) circle (4pt);
    \end{tikzpicture}} } \times \frac{1}{H_{\begin{tikzpicture}[scale=0.8]
        \coordinate (A) at (0,0);
        \coordinate (B) at (1/2,0);
        \draw[thick] (A) -- (B);
        \fill[black] (A) circle (2pt);
        \fill[black] (B) circle (2pt);
        \draw[thick] (B) to[out=90,in=90] (1,0) to[out=90+180,in=90+180] (B);
        \draw[black,thick] (B) circle (4pt);
    \end{tikzpicture}}}.
\end{align}

\subsection{Decorated amplitubes}
In this section we show how \eqref{eq:edge_expand} reproduces a recently conjectured formula for the wavefunction coefficient motivated by partial fractions \cite{Fevola:2024nzj}, see formula (15) therein. In order to do so we simply substitute the expressions for the amplitubes decomposed into directed graphs i.e.~\eqref{eq:amp_or} into \eqref{eq:edge_expand} to arrive at 
\begin{align}
\Psi_G = \sum_{{\bf e} \subset E_G} (-1)^{|{\bf e}|}  \prod_{b \in {\bf b_e}} \sum_{b^\circ \in b^{\text{dir}}} A_{b^{\circ}}.
\label{eq:main}
\end{align}
Each term appearing on the right hand side of \eqref{eq:main}, which we refer to as decorated amplitubes, can be labelled by decorating every edge of the graph $G$ by either a broken or directed edge depicted as $\begin{tikzpicture}[scale=1]
        \coordinate (A) at (0,0);
        \coordinate (B) at (1/2,0);
        \coordinate (C) at (1,0);
        \coordinate (D) at (3/2,0);
        \draw[thick,dotted] (A) -- (B);
        \fill[black] (A) circle (2pt);
        \fill[black] (B) circle (2pt);
    \end{tikzpicture}$, $\raisebox{-0.11cm}{\begin{tikzpicture}[scale=1]
        \coordinate (A) at (0,0);
        \coordinate (B) at (1/2,0);
        \coordinate (C) at (1,0);
        \coordinate (D) at (3/2,0);
        \draw[thick] (A) --  node {\al} (B);
        \fill[black] (A) circle (2pt);
        \fill[black] (B) circle (2pt);
    \end{tikzpicture}}$ or $\raisebox{-0.11cm}{\begin{tikzpicture}[scale=1]
        \coordinate (A) at (0,0);
        \coordinate (B) at (1/2,0);
        \coordinate (C) at (1,0);
        \coordinate (D) at (3/2,0);
        \draw[thick] (A) --  node {\ar} (B);
        \fill[black] (A) circle (2pt);
        \fill[black] (B) circle (2pt);
    \end{tikzpicture}}$. Again, by convention we treat any unbroken loop edge as un-oriented. To demonstrate this formula, and finish our running examples, consider the path graph on three vertices whose expansion takes the following form
\begin{align}
\Psi_{\begin{tikzpicture}[scale=0.8]
        \coordinate (A) at (0,0);
        \coordinate (B) at (1/2,0);
        \coordinate (C) at (1,0);
        \coordinate (D) at (3/2,0);
        \draw[thick] (A) -- (B) -- (C) ;
        \fill[black] (A) circle (2pt);
        \fill[black] (B) circle (2pt);
        \fill[black] (C) circle (2pt);
    \end{tikzpicture}}&=\underbrace{A_{\begin{tikzpicture}[scale=0.8]
        \coordinate (A) at (0,0);
        \coordinate (B) at (1/2,0);
        \coordinate (C) at (1,0);
        \coordinate (D) at (3/2,0);
        \draw[thick] (A) -- (B) -- (C) ;
        \draw (A)-- node {\ar} (B);
        \draw (B)-- node {\ar} (C);
        \fill[black] (A) circle (2pt);
        \fill[black] (B) circle (2pt);
        \fill[black] (C) circle (2pt);
    \end{tikzpicture}}+A_{\begin{tikzpicture}[scale=0.8]
        \coordinate (A) at (0,0);
        \coordinate (B) at (1/2,0);
        \coordinate (C) at (1,0);
        \coordinate (D) at (3/2,0);
        \draw[thick] (A) -- (B) -- (C) ;
        \draw (A)-- node {\ar} (B);
        \draw (B)-- node {\al} (C);
        \fill[black] (A) circle (2pt);
        \fill[black] (B) circle (2pt);
        \fill[black] (C) circle (2pt);
    \end{tikzpicture}}+A_{\begin{tikzpicture}[scale=0.8]
        \coordinate (A) at (0,0);
        \coordinate (B) at (1/2,0);
        \coordinate (C) at (1,0);
        \coordinate (D) at (3/2,0);
        \draw[thick] (A) -- (B) -- (C) ;
        \draw (A)-- node {\al} (B);
        \draw (B)-- node {\ar} (C);
        \fill[black] (A) circle (2pt);
        \fill[black] (B) circle (2pt);
        \fill[black] (C) circle (2pt);
    \end{tikzpicture}} +A_{\begin{tikzpicture}[scale=0.8]
        \coordinate (A) at (0,0);
        \coordinate (B) at (1/2,0);
        \coordinate (C) at (1,0);
        \coordinate (D) at (3/2,0);
        \draw[thick] (A) -- (B) -- (C) ;
        \draw (A)-- node {\al} (B);
        \draw (B)-- node {\al} (C);
        \fill[black] (A) circle (2pt);
        \fill[black] (B) circle (2pt);
        \fill[black] (C) circle (2pt);
    \end{tikzpicture}}}_{A_{\begin{tikzpicture}[scale=0.8]
        \coordinate (A) at (0,0);
        \coordinate (B) at (1/2,0);
        \coordinate (C) at (1,0);
        \coordinate (D) at (3/2,0);
        \draw[thick] (A) -- (B) -- (C) ;
        \fill[black] (A) circle (2pt);
        \fill[black] (B) circle (2pt);
        \fill[black] (C) circle (2pt);
    \end{tikzpicture}}}
    -  \underbrace{A_{\begin{tikzpicture}[scale=0.8]
        \coordinate (A) at (0,0);
        \coordinate (B) at (1/2,0);
        \coordinate (C) at (1,0);
        \coordinate (D) at (3/2,0);
        \draw[thick] (A) -- (B) ;
        \draw[thick,dotted] (B) -- (C) ;
        \draw (A)-- node {\ar} (B);
        \fill[black] (A) circle (2pt);
        \fill[black] (B) circle (2pt);
        \fill[black] (C) circle (2pt);
    \end{tikzpicture}}-A_{\begin{tikzpicture}[scale=0.8]
        \coordinate (A) at (0,0);
        \coordinate (B) at (1/2,0);
        \coordinate (C) at (1,0);
        \coordinate (D) at (3/2,0);
        \draw[thick] (A) -- (B) ;
        \draw[thick,dotted] (B) -- (C) ;
        \draw (A)-- node {\al} (B);
        \fill[black] (A) circle (2pt);
        \fill[black] (B) circle (2pt);
        \fill[black] (C) circle (2pt);
    \end{tikzpicture}}}_{-A_{\begin{tikzpicture}[scale=0.8]
        \coordinate (A) at (0,0);
        \coordinate (B) at (1/2,0);
        \coordinate (C) at (1,0);
        \coordinate (D) at (3/2,0);
        \draw[thick] (A) -- (B) ;
        \draw[thick,dotted] (B) -- (C) ;
        \fill[black] (A) circle (2pt);
        \fill[black] (B) circle (2pt);
        \fill[black] (C) circle (2pt);
    \end{tikzpicture}}} \notag \\
    &- \underbrace{A_{\begin{tikzpicture}[scale=0.8]
        \coordinate (A) at (0,0);
        \coordinate (B) at (1/2,0);
        \coordinate (C) at (1,0);
        \coordinate (D) at (3/2,0);
        \draw[thick] (B) -- (C) ;
        \draw[thick,dotted] (A) -- (B) ;
        \draw (B)-- node {\ar} (C);
        \fill[black] (A) circle (2pt);
        \fill[black] (B) circle (2pt);
        \fill[black] (C) circle (2pt);
    \end{tikzpicture}}-A_{\begin{tikzpicture}[scale=0.8]
        \coordinate (A) at (0,0);
        \coordinate (B) at (1/2,0);
        \coordinate (C) at (1,0);
        \coordinate (D) at (3/2,0);
        \draw[thick] (B) -- (C) ;
        \draw[thick,dotted] (A) -- (B) ;
        \draw (B)-- node {\al} (C);
        \fill[black] (A) circle (2pt);
        \fill[black] (B) circle (2pt);
        \fill[black] (C) circle (2pt);
    \end{tikzpicture}}}_{-A_{\begin{tikzpicture}[scale=0.8]
        \coordinate (A) at (0,0);
        \coordinate (B) at (1/2,0);
        \coordinate (C) at (1,0);
        \coordinate (D) at (3/2,0);
        \draw[thick,dotted] (A) -- (B) ;
        \draw[thick] (B) -- (C) ;
        \fill[black] (A) circle (2pt);
        \fill[black] (B) circle (2pt);
        \fill[black] (C) circle (2pt);
    \end{tikzpicture}}}+A_{\begin{tikzpicture}[scale=0.8]
        \coordinate (A) at (0,0);
        \coordinate (B) at (1/2,0);
        \coordinate (C) at (1,0);
        \coordinate (D) at (3/2,0);
        \draw[thick,dotted] (A) -- (B) ;
        \draw[thick,dotted] (B) -- (C) ;
        \fill[black] (A) circle (2pt);
        \fill[black] (B) circle (2pt);
        \fill[black] (C) circle (2pt);
    \end{tikzpicture}}.
\end{align}
The decorated graphs containing both directed and broken edges are given explicitly by
\begin{align}
A_{\begin{tikzpicture}[scale=0.8]
        \coordinate (A) at (0,0);
        \coordinate (B) at (1/2,0);
        \coordinate (C) at (1,0);
        \coordinate (D) at (3/2,0);
        \draw[thick,dotted] (B) -- (C) ;
        \draw[thick] (A)-- node {\ar} (B);
        \fill[black] (A) circle (2pt);
        \fill[black] (B) circle (2pt);
        \fill[black] (C) circle (2pt);
    \end{tikzpicture}}& = \frac{1}{H_{ \begin{tikzpicture}[scale=0.8]
        \coordinate (A) at (0,0);
        \coordinate (B) at (1/2,0);
        \coordinate (C) at (1,0);
        \coordinate (D) at (3/2,0);
        \draw[thick, black] (0.25,0) ellipse (0.41cm and 0.19cm);
        \draw[thick] (A) -- (B) -- (C);
        \draw[black,thick] (A) circle (3pt);
        \draw[black,thick] (C) circle (3pt);
        \fill[black] (A) circle (2pt);
        \fill[black] (B) circle (2pt);
        \fill[black] (C) circle (2pt);
    \end{tikzpicture} }}, \quad A_{\begin{tikzpicture}[scale=0.8]
        \coordinate (A) at (0,0);
        \coordinate (B) at (1/2,0);
        \coordinate (C) at (1,0);
        \coordinate (D) at (3/2,0);
        \draw[thick,dotted] (B) -- (C) ;
        \draw[thick] (A)-- node {\al} (B);
        \fill[black] (A) circle (2pt);
        \fill[black] (B) circle (2pt);
        \fill[black] (C) circle (2pt);
    \end{tikzpicture}}= \frac{1}{H_{ \begin{tikzpicture}[scale=0.8]
        \coordinate (A) at (0,0);
        \coordinate (B) at (1/2,0);
        \coordinate (C) at (1,0);
        \coordinate (D) at (3/2,0);
        \draw[thick, black] (0.25,0) ellipse (0.41cm and 0.19cm);
        \draw[thick] (A) -- (B) -- (C);
        \draw[black,thick] (B) circle (3pt);
        \draw[black,thick] (C) circle (3pt);
        \fill[black] (A) circle (2pt);
        \fill[black] (B) circle (2pt);
        \fill[black] (C) circle (2pt);
    \end{tikzpicture} }},  \quad 
A_{\begin{tikzpicture}[scale=0.8]
        \coordinate (A) at (0,0);
        \coordinate (B) at (1/2,0);
        \coordinate (C) at (1,0);
        \coordinate (D) at (3/2,0);
        \draw[thick] (B) -- (C) ;
        \draw[thick,dotted] (A) -- (B) ;
        \draw (B)-- node {\ar} (C);
        \fill[black] (A) circle (2pt);
        \fill[black] (B) circle (2pt);
        \fill[black] (C) circle (2pt);
    \end{tikzpicture}}= \frac{1}{H_{ \begin{tikzpicture}[scale=0.8]
        \coordinate (A) at (0,0);
        \coordinate (B) at (1/2,0);
        \coordinate (C) at (1,0);
        \coordinate (D) at (3/2,0);
        \draw[thick, black] (0.75,0) ellipse (0.41cm and 0.19cm);
        \draw[thick] (A) -- (B) -- (C);
        \draw[black,thick] (A) circle (3pt);
        \draw[black,thick] (B) circle (3pt);
        \fill[black] (A) circle (2pt);
        \fill[black] (B) circle (2pt);
        \fill[black] (C) circle (2pt);
    \end{tikzpicture} }}, \quad
     A_{\begin{tikzpicture}[scale=0.8]
        \coordinate (A) at (0,0);
        \coordinate (B) at (1/2,0);
        \coordinate (C) at (1,0);
        \coordinate (D) at (3/2,0);
        \draw[thick] (B) -- (C) ;
        \draw[thick,dotted] (A) -- (B) ;
        \draw (B)-- node {\al} (C);
        \fill[black] (A) circle (2pt);
        \fill[black] (B) circle (2pt);
        \fill[black] (C) circle (2pt);
    \end{tikzpicture}}= \frac{1}{H_{ \begin{tikzpicture}[scale=0.8]
        \coordinate (A) at (0,0);
        \coordinate (B) at (1/2,0);
        \coordinate (C) at (1,0);
        \coordinate (D) at (3/2,0);
        \draw[thick, black] (0.75,0) ellipse (0.41cm and 0.19cm);
        \draw[thick] (A) -- (B) -- (C);
        \draw[black,thick] (A) circle (3pt);
        \draw[black,thick] (C) circle (3pt);
        \fill[black] (A) circle (2pt);
        \fill[black] (B) circle (2pt);
        \fill[black] (C) circle (2pt);
    \end{tikzpicture} }}. \notag 
\end{align}
Next, consider the two-cycle which has the following expansion
\begin{align}
\Psi_{
\begin{tikzpicture}[scale=0.6]
\fill[black] (0,-1) circle (2pt);
\fill[black] (1,-1) circle (2pt);
\draw[thick] (0,-1) to[out=90,in=90] (1,-1);
\draw[thick] (0,-1) to[out=-90,in=180+90] (1,-1);
\end{tikzpicture}}=&\underbrace{A_{
\begin{tikzpicture}[scale=0.6]
\fill[black] (0,-1) circle (2pt);
\fill[black] (1,-1) circle (2pt);
\draw[thick] (0,-1) to[out=90,in=90] node {\ar} (1,-1);
\draw[thick] (0,-1) to[out=-90,in=180+90] node {\ar} (1,-1);
\end{tikzpicture}}+A_{
\begin{tikzpicture}[scale=0.6]
\fill[black] (0,-1) circle (2pt);
\fill[black] (1,-1) circle (2pt);
\draw[thick] (0,-1) to[out=90,in=90] node {\al} (1,-1);
\draw[thick] (0,-1) to[out=-90,in=180+90] node {\al} (1,-1);
\end{tikzpicture}}}_{A_{
\begin{tikzpicture}[scale=0.6]
\fill[black] (0,-1) circle (2pt);
\fill[black] (1,-1) circle (2pt);
\draw[thick] (0,-1) to[out=90,in=90] (1,-1);
\draw[thick] (0,-1) to[out=-90,in=180+90]  (1,-1);
\end{tikzpicture}}}-\underbrace{A_{
\begin{tikzpicture}[scale=0.6]
\fill[black] (0,-1) circle (2pt);
\fill[black] (1,-1) circle (2pt);
\draw[thick] (0,-1) to[out=90,in=90] node {\ar} (1,-1);
\draw[thick,dotted] (0,-1) to[out=-90,in=180+90] (1,-1);
\end{tikzpicture}}-A_{
\begin{tikzpicture}[scale=0.6]
\fill[black] (0,-1) circle (2pt);
\fill[black] (1,-1) circle (2pt);
\draw[thick] (0,-1) to[out=90,in=90] node {\al} (1,-1);
\draw[thick,dotted] (0,-1) to[out=-90,in=180+90] (1,-1);
\end{tikzpicture}}}_{-A_{
\begin{tikzpicture}[scale=0.6]
\fill[black] (0,-1) circle (2pt);
\fill[black] (1,-1) circle (2pt);
\draw[thick] (0,-1) to[out=90,in=90] (1,-1);
\draw[thick,dotted] (0,-1) to[out=-90,in=180+90]  (1,-1);
\end{tikzpicture}}}
-\underbrace{A_{
\begin{tikzpicture}[scale=0.6]
\fill[black] (0,-1) circle (2pt);
\fill[black] (1,-1) circle (2pt);
\draw[thick,dotted] (0,-1) to[out=90,in=90]  (1,-1);
\draw[thick] (0,-1) to[out=-90,in=180+90] node {\ar} (1,-1);
\end{tikzpicture}}-A_{
\begin{tikzpicture}[scale=0.6]
\fill[black] (0,-1) circle (2pt);
\fill[black] (1,-1) circle (2pt);
\draw[thick,dotted] (0,-1) to[out=90,in=90]  (1,-1);
\draw[thick] (0,-1) to[out=-90,in=180+90] node {\al} (1,-1);
\end{tikzpicture}}}_{-A_{
\begin{tikzpicture}[scale=0.6]
\fill[black] (0,-1) circle (2pt);
\fill[black] (1,-1) circle (2pt);
\draw[thick,dotted] (0,-1) to[out=90,in=90] (1,-1);
\draw[thick] (0,-1) to[out=-90,in=180+90]  (1,-1);
\end{tikzpicture}}}+A_{
\begin{tikzpicture}[scale=0.6]
\fill[black] (0,-1) circle (2pt);
\fill[black] (1,-1) circle (2pt);
\draw[thick,dotted] (0,-1) to[out=90,in=90]  (1,-1);
\draw[thick,dotted] (0,-1) to[out=-90,in=180+90] (1,-1);
\end{tikzpicture}},
\end{align}
where the graphs with both broken and directed edges are given by 
\begin{align}
A_{
\begin{tikzpicture}[scale=0.6]
\fill[black] (0,-1) circle (2pt);
\fill[black] (1,-1) circle (2pt);
\draw[thick] (0,-1) to[out=90,in=90] node {\ar} (1,-1);
\draw[thick,dotted] (0,-1) to[out=-90,in=180+90] (1,-1);
\end{tikzpicture}} &= \frac{1}{H_{
\begin{tikzpicture}[scale=0.6]
\fill[black] (0,-1) circle (2pt);
\fill[black] (1,-1) circle (2pt);
\draw[black,thick] (0,-1) circle (3.5pt);
\draw[thick] (0,-1) to[out=90,in=90] (1,-1);
\draw[thick] (0,-1) to[out=-90,in=180+90] (1,-1);
\draw[thick] (0-0.2,-1) to[out=90,in=90] (1+0.2,-1);
\draw[thick] (0+0.2,-1) to[out=90,in=90] (1-0.2,-1);
\draw[thick] (1-0.2,-1) to[out=90+180,in=180] (1,-1.2) to[out=0,in=90+180] (1+0.2,-1);
\draw[thick] (0-0.2,-1) to[out=90+180,in=180] (0,-1.2) to[out=0,in=90+180] (0+0.2,-1);
\end{tikzpicture}}}, \quad \quad A_{
\begin{tikzpicture}[scale=0.6]
\fill[black] (0,-1) circle (2pt);
\fill[black] (1,-1) circle (2pt);
\draw[thick] (0,-1) to[out=90,in=90] node {\al} (1,-1);
\draw[thick,dotted] (0,-1) to[out=-90,in=180+90] (1,-1);
\end{tikzpicture}} = \frac{1}{H_{
\begin{tikzpicture}[scale=0.6]
\fill[black] (0,-1) circle (2pt);
\fill[black] (1,-1) circle (2pt);
\draw[black,thick] (1,-1) circle (3.5pt);
\draw[thick] (0,-1) to[out=90,in=90] (1,-1);
\draw[thick] (0,-1) to[out=-90,in=180+90] (1,-1);
\draw[thick] (0-0.2,-1) to[out=90,in=90] (1+0.2,-1);
\draw[thick] (0+0.2,-1) to[out=90,in=90] (1-0.2,-1);
\draw[thick] (1-0.2,-1) to[out=90+180,in=180] (1,-1.2) to[out=0,in=90+180] (1+0.2,-1);
\draw[thick] (0-0.2,-1) to[out=90+180,in=180] (0,-1.2) to[out=0,in=90+180] (0+0.2,-1);
\end{tikzpicture}}}, \quad \quad 
A_{
\begin{tikzpicture}[scale=0.6]
\fill[black] (0,-1) circle (2pt);
\fill[black] (1,-1) circle (2pt);
\draw[thick,dotted] (0,-1) to[out=90,in=90]  (1,-1);
\draw[thick] (0,-1) to[out=-90,in=180+90] node {\ar} (1,-1);
\end{tikzpicture}}=  \frac{1}{H_{
\begin{tikzpicture}[scale=0.6]
\fill[black] (0,-1) circle (2pt);
\fill[black] (1,-1) circle (2pt);
\draw[black,thick] (0,-1) circle (3.5pt);
\draw[thick] (0,-1) to[out=90,in=90] (1,-1);
\draw[thick] (0,-1) to[out=-90,in=180+90] (1,-1);
\draw[thick] (0-0.2,-1) to[out=-90,in=180+90] (1+0.2,-1);
\draw[thick] (0+0.2,-1) to[out=-90,in=180+90] (1-0.2,-1);
\draw[thick] (1-0.2,-1) to[out=90,in=180] (1,-0.8) to[out=0,in=90] (1+0.2,-1);
\draw[thick] (0-0.2,-1) to[out=90,in=180] (0,-0.8) to[out=0,in=90] (0+0.2,-1);
\end{tikzpicture}}}, \quad \quad 
 A_{
\begin{tikzpicture}[scale=0.6]
\fill[black] (0,-1) circle (2pt);
\fill[black] (1,-1) circle (2pt);
\draw[thick,dotted] (0,-1) to[out=90,in=90]  (1,-1);
\draw[thick] (0,-1) to[out=-90,in=180+90] node {\al} (1,-1);
\end{tikzpicture}}= \frac{1}{H_{
\begin{tikzpicture}[scale=0.6]
\fill[black] (0,-1) circle (2pt);
\fill[black] (1,-1) circle (2pt);
\draw[black,thick] (1,-1) circle (3.5pt);
\draw[thick] (0,-1) to[out=90,in=90] (1,-1);
\draw[thick] (0,-1) to[out=-90,in=180+90] (1,-1);
\draw[thick] (0-0.2,-1) to[out=-90,in=180+90] (1+0.2,-1);
\draw[thick] (0+0.2,-1) to[out=-90,in=180+90] (1-0.2,-1);
\draw[thick] (1-0.2,-1) to[out=90,in=180] (1,-0.8) to[out=0,in=90] (1+0.2,-1);
\draw[thick] (0-0.2,-1) to[out=90,in=180] (0,-0.8) to[out=0,in=90] (0+0.2,-1);
\end{tikzpicture}}}.
\end{align}
Finally, we have the graph containing a loop which has the following expansion
\begin{align}
\Psi_{\begin{tikzpicture}[scale=0.8]
        \coordinate (A) at (0,0);
        \coordinate (B) at (1/2,0);
        \draw[thick] (A) -- (B);
        \fill[black] (A) circle (2pt);
        \fill[black] (B) circle (2pt);
        \draw[thick] (B) to[out=90,in=90] (1,0) to[out=90+180,in=90+180] (B);
    \end{tikzpicture}} = \underbrace{A_{\begin{tikzpicture}[scale=0.8]
        \coordinate (A) at (0,0);
        \coordinate (B) at (1/2,0);
        \draw[thick] (A) -- node {\ar} (B);
        \fill[black] (A) circle (2pt);
        \fill[black] (B) circle (2pt);
        \draw[thick] (B) to[out=90,in=90] (1,0) to[out=90+180,in=90+180] (B);
    \end{tikzpicture}} +A_{\begin{tikzpicture}[scale=0.8]
        \coordinate (A) at (0,0);
        \coordinate (B) at (1/2,0);
        \draw[thick] (A) -- node {\al} (B);
        \fill[black] (A) circle (2pt);
        \fill[black] (B) circle (2pt);
        \draw[thick] (B) to[out=90,in=90] (1,0) to[out=90+180,in=90+180] (B);
    \end{tikzpicture}}}_{A_{\begin{tikzpicture}[scale=0.8]
        \coordinate (A) at (0,0);
        \coordinate (B) at (1/2,0);
        \draw[thick] (A) --  (B);
        \fill[black] (A) circle (2pt);
        \fill[black] (B) circle (2pt);
        \draw[thick] (B) to[out=90,in=90] (1,0) to[out=90+180,in=90+180] (B);
    \end{tikzpicture}}} -\underbrace{A_{\begin{tikzpicture}[scale=0.8]
        \coordinate (A) at (0,0);
        \coordinate (B) at (1/2,0);
        \draw[thick] (A) -- node {\ar} (B);
        \fill[black] (A) circle (2pt);
        \fill[black] (B) circle (2pt);
        \draw[thick,dotted] (B) to[out=90,in=90] (1,0) to[out=90+180,in=90+180] (B);
    \end{tikzpicture}} -A_{\begin{tikzpicture}[scale=0.8]
        \coordinate (A) at (0,0);
        \coordinate (B) at (1/2,0);
        \draw[thick] (A) -- node {\al} (B);
        \fill[black] (A) circle (2pt);
        \fill[black] (B) circle (2pt);
        \draw[thick,dotted] (B) to[out=90,in=90] (1,0) to[out=90+180,in=90+180] (B);
    \end{tikzpicture}}}_{-A_{\begin{tikzpicture}[scale=0.8]
        \coordinate (A) at (0,0);
        \coordinate (B) at (1/2,0);
        \draw[thick] (A) -- (B);
        \fill[black] (A) circle (2pt);
        \fill[black] (B) circle (2pt);
        \draw[thick,dotted] (B) to[out=90,in=90] (1,0) to[out=90+180,in=90+180] (B);
    \end{tikzpicture}}} -A_{\begin{tikzpicture}[scale=0.8]
        \coordinate (A) at (0,0);
        \coordinate (B) at (1/2,0);e
        \draw[thick,dotted] (A) -- (B);
        \fill[black] (A) circle (2pt);
        \fill[black] (B) circle (2pt);
        \draw[thick] (B) to[out=90,in=90] (1,0) to[out=90+180,in=90+180] (B);
    \end{tikzpicture}} +A_{\begin{tikzpicture}[scale=0.8]
        \coordinate (A) at (0,0);
        \coordinate (B) at (1/2,0);
        \draw[thick,dotted] (A) -- (B);
        \fill[black] (A) circle (2pt);
        \fill[black] (B) circle (2pt);
        \draw[thick,dotted] (B) to[out=90,in=90] (1,0) to[out=90+180,in=90+180] (B);
    \end{tikzpicture}},
\end{align}
where the graphs with both broken and directed edges are given by 
\begin{align}
A_{\begin{tikzpicture}[scale=0.8]
        \coordinate (A) at (0,0);
        \coordinate (B) at (1/2,0);
        \draw[thick] (A) -- node {\ar} (B);
        \fill[black] (A) circle (2pt);
        \fill[black] (B) circle (2pt);
        \draw[thick,dotted] (B) to[out=90,in=90] (1,0) to[out=90+180,in=90+180] (B);
    \end{tikzpicture}} =  \frac{1}{H_{\begin{tikzpicture}[scale=0.8]
        \coordinate (A) at (0,0);
        \coordinate (B) at (1/2,0);
        \draw[thick] (A) -- (B);
        \fill[black] (A) circle (2pt);
        \fill[black] (B) circle (2pt);
        \draw[thick] (B) to[out=90,in=90] (1,0) to[out=90+180,in=90+180] (B);
        \draw[thick, black] (0.25,0) ellipse (0.5cm and 0.22cm);
        \draw[black,thick] (A) circle (4pt);
    \end{tikzpicture}} },\quad \quad A_{\begin{tikzpicture}[scale=0.8]
        \coordinate (A) at (0,0);
        \coordinate (B) at (1/2,0);
        \draw[thick] (A) -- node {\al} (B);
        \fill[black] (A) circle (2pt);
        \fill[black] (B) circle (2pt);
        \draw[thick,dotted] (B) to[out=90,in=90] (1,0) to[out=90+180,in=90+180] (B);
    \end{tikzpicture}}=\frac{1}{H_{\begin{tikzpicture}[scale=0.8]
        \coordinate (A) at (0,0);
        \coordinate (B) at (1/2,0);
        \draw[thick] (A) -- (B);
        \fill[black] (A) circle (2pt);
        \fill[black] (B) circle (2pt);
        \draw[thick] (B) to[out=90,in=90] (1,0) to[out=90+180,in=90+180] (B);
        \draw[thick, black] (0.25,0) ellipse (0.5cm and 0.22cm);
        \draw[black,thick] (B) circle (4pt);
    \end{tikzpicture}} }.
\end{align}
%========================================================================================
%========================================================================================
\section{Cut tubings}
\label{sec:cut_tub}
The result for the wavefunction coefficients presented in \eqref{eq:edge_expand} suggests to us a new definition of tubing which makes use of both the binary and unary tubes, we refer to these as {\it cut} tubings, see also \cite{De:2024zic}. A cut tubing ${\bf b_{e,(u_1,\ldots, u_k)}}$ can be constructed from the following data:
\begin{itemize}
\item a subset of cut edges ${\bf e} \subset E_G$ or equivalently the corresponding partition tube ${\bf b_e}=(b_1,\ldots,b_k)$,
\item a $u$-tubing ${\bf u}_i$ (not necessarily maximal) for each $b_i \in {\bf b_e}$.
\end{itemize}
Given this data the corresponding cut tubing is defined as ${\bf b_{e,(u_1,\ldots, u_k)}}= \cup_{i} {\bf u}_i$. Furthermore, each cut tubing ${\bf b_{e,u_1,\ldots, u_k}}$ induces a {\it decorated orientation} of the graph as follows:
\begin{itemize}
\item each edge $e \in {\bf e}$ is decorated with a broken edge $\begin{tikzpicture}[scale=1.1]
        \coordinate (A) at (0,0);
        \coordinate (B) at (1/2,0);
        \coordinate (C) at (1,0);
        \coordinate (D) at (3/2,0);
        \draw[thick,dotted] (A) -- (B);
        \fill[black] (A) circle (2pt);
        \fill[black] (B) circle (2pt);
    \end{tikzpicture}$,
 \item all edges in $E_G \setminus {\bf e}$ cut by a tube are oriented following the rule \eqref{eq:or_rule},
 \item all remaining edges are decorated with a solid edge $\begin{tikzpicture}[scale=1.1]
        \coordinate (A) at (0,0);
        \coordinate (B) at (1/2,0);
        \coordinate (C) at (1,0);
        \coordinate (D) at (3/2,0);
        \draw[thick] (A) -- (B);
        \fill[black] (A) circle (2pt);
        \fill[black] (B) circle (2pt);
    \end{tikzpicture}$.
 
\end{itemize}
As an example the cut tubings and corresponding decorated orientations of the path graph on three vertices are given by
\begin{center}
\begin{tabular}{cccc}
$\raisebox{-0.16cm}{\begin{tikzpicture}[scale=1]
        \coordinate (A) at (0,0);
        \coordinate (B) at (1/2,0);
        \coordinate (C) at (1,0);
        \draw[thick, black] (0.5,0) ellipse (0.75cm and 0.25cm);
        \draw[thick] (A) -- (B) -- (C);
        \fill[black] (A) circle (2pt);
        \fill[black] (B) circle (2pt);
        \fill[black] (C) circle (2pt);
    \end{tikzpicture}} \leftrightarrow \raisebox{0.03cm}{\begin{tikzpicture}[scale=1]
        \coordinate (A) at (0,0);
        \coordinate (B) at (1/2,0);
        \coordinate (C) at (1,0);
        \draw[thick] (A) -- (B) -- (C);
        \fill[black] (A) circle (2pt);
        \fill[black] (B) circle (2pt);
        \fill[black] (C) circle (2pt);
    \end{tikzpicture}}$, \hfill & $\raisebox{-0.1cm}{\begin{tikzpicture}[scale=1]
        \coordinate (A) at (0,0);
        \coordinate (B) at (1/2,0);
        \coordinate (C) at (1,0);
        \draw[thick, black] (0.25,0) ellipse (0.41cm and 0.19cm);
        \draw[thick] (A) -- (B) -- (C);
        \draw[black,thick] (C) circle (3pt);
        \fill[black] (A) circle (2pt);
        \fill[black] (B) circle (2pt);
        \fill[black] (C) circle (2pt);
    \end{tikzpicture}} \leftrightarrow \raisebox{0.03cm}{\begin{tikzpicture}[scale=1]
        \coordinate (A) at (0,0);
        \coordinate (B) at (1/2,0);
        \coordinate (C) at (1,0);
        \draw[thick] (A) -- (B);
        \draw[thick,dotted] (B) -- (C);
        \fill[black] (A) circle (2pt);
        \fill[black] (B) circle (2pt);
        \fill[black] (C) circle (2pt);
    \end{tikzpicture}}$, \hfill& $\raisebox{-0.1cm}{\begin{tikzpicture}[scale=1]
        \coordinate (A) at (0,0);
        \coordinate (B) at (1/2,0);
        \coordinate (C) at (1,0);
        \draw[thick, black] (0.75,0) ellipse (0.41cm and 0.19cm);
        \draw[thick] (A) -- (B) -- (C);
        \draw[black,thick] (A) circle (3pt);
        \fill[black] (A) circle (2pt);
        \fill[black] (B) circle (2pt);
        \fill[black] (C) circle (2pt);
    \end{tikzpicture}} \leftrightarrow \raisebox{0.03cm}{\begin{tikzpicture}[scale=1]
        \coordinate (A) at (0,0);
        \coordinate (B) at (1/2,0);
        \coordinate (C) at (1,0);
        \draw[thick,dotted] (A) -- (B);
        \draw[thick] (B) -- (C);
        \fill[black] (A) circle (2pt);
        \fill[black] (B) circle (2pt);
        \fill[black] (C) circle (2pt);
    \end{tikzpicture}}$, \hfill& $\raisebox{-0.02cm}{\begin{tikzpicture}[scale=1]
        \coordinate (A) at (0,0);
        \coordinate (B) at (1/2,0);
        \coordinate (C) at (1,0);
        \draw[thick] (A) -- (B) -- (C);
        \draw[black,thick] (A) circle (3pt);
        \draw[black,thick] (B) circle (3pt);
        \draw[black,thick] (C) circle (3pt);
        \fill[black] (A) circle (2pt);
        \fill[black] (B) circle (2pt);
        \fill[black] (C) circle (2pt);
    \end{tikzpicture}} \leftrightarrow \raisebox{0.03cm}{\begin{tikzpicture}[scale=1]
        \coordinate (A) at (0,0);
        \coordinate (B) at (1/2,0);
        \coordinate (C) at (1,0);
        \draw[thick,dotted] (A) -- (B) -- (C);
        \fill[black] (A) circle (2pt);
        \fill[black] (B) circle (2pt);
        \fill[black] (C) circle (2pt);
    \end{tikzpicture}}$,\hfill  \\[0.8em]
$\raisebox{-0.16cm}{\begin{tikzpicture}[scale=1]
        \coordinate (A) at (0,0);
        \coordinate (B) at (1/2,0);
        \coordinate (C) at (1,0);
        \draw[thick, black] (0.5,0) ellipse (0.75cm and 0.25cm);
        \draw[thick] (A) -- (B) -- (C);
        \fill[black] (A) circle (2pt);
        \fill[black] (B) circle (2pt);
        \fill[black] (C) circle (2pt);
        \draw[black,thick] (A) circle (4pt);
    \end{tikzpicture}} \leftrightarrow \raisebox{-0.07cm}{\begin{tikzpicture}[scale=1]
        \coordinate (A) at (0,0);
        \coordinate (B) at (1/2,0);
        \coordinate (C) at (1,0);
        \draw[thick] (A) -- node {\ar} (B) -- (C);
        \fill[black] (A) circle (2pt);
        \fill[black] (B) circle (2pt);
        \fill[black] (C) circle (2pt);
    \end{tikzpicture}}$, \hfill & $\raisebox{-0.16cm}{\begin{tikzpicture}[scale=1]
        \coordinate (A) at (0,0);
        \coordinate (B) at (1/2,0);
        \coordinate (C) at (1,0);
        \draw[thick, black] (0.5,0) ellipse (0.75cm and 0.25cm);
        \draw[thick] (A) -- (B) -- (C);
        \fill[black] (A) circle (2pt);
        \fill[black] (B) circle (2pt);
        \fill[black] (C) circle (2pt);
        \draw[black,thick] (B) circle (4pt);
    \end{tikzpicture}} \leftrightarrow \raisebox{-0.07cm}{\begin{tikzpicture}[scale=1]
        \coordinate (A) at (0,0);
        \coordinate (B) at (1/2,0);
        \coordinate (C) at (1,0);
        \draw[thick] (A) -- node {\al} (B) -- node {\ar} (C);
        \fill[black] (A) circle (2pt);
        \fill[black] (B) circle (2pt);
        \fill[black] (C) circle (2pt);
    \end{tikzpicture}}$\hfill & $\raisebox{-0.16cm}{\begin{tikzpicture}[scale=1]
        \coordinate (A) at (0,0);
        \coordinate (B) at (1/2,0);
        \coordinate (C) at (1,0);
        \draw[thick, black] (0.5,0) ellipse (0.75cm and 0.25cm);
        \draw[thick] (A) -- (B) -- (C);
        \fill[black] (A) circle (2pt);
        \fill[black] (B) circle (2pt);
        \fill[black] (C) circle (2pt);
        \draw[black,thick] (C) circle (4pt);
    \end{tikzpicture}} \leftrightarrow \raisebox{-0.07cm}{\begin{tikzpicture}[scale=1]
        \coordinate (A) at (0,0);
        \coordinate (B) at (1/2,0);
        \coordinate (C) at (1,0);
        \draw[thick] (A) -- (B) -- node {\al} (C);
        \fill[black] (A) circle (2pt);
        \fill[black] (B) circle (2pt);
        \fill[black] (C) circle (2pt);
    \end{tikzpicture}}$ \hfill & $\raisebox{-0.16cm}{\begin{tikzpicture}[scale=1]
        \coordinate (A) at (0,0);
        \coordinate (B) at (1/2,0);
        \coordinate (C) at (1,0);
        \draw[thick, black] (0.5,0) ellipse (0.75cm and 0.25cm);
         \draw[thick, black] (0.25,0) ellipse (0.41cm and 0.19cm);
        \draw[thick] (A) -- (B) -- (C);
        \fill[black] (A) circle (2pt);
        \fill[black] (B) circle (2pt);
        \fill[black] (C) circle (2pt);
    \end{tikzpicture}} \leftrightarrow \raisebox{-0.07cm}{\begin{tikzpicture}[scale=1]
        \coordinate (A) at (0,0);
        \coordinate (B) at (1/2,0);
        \coordinate (C) at (1,0);
        \draw[thick] (A) -- (B) -- node {\ar} (C);
        \fill[black] (A) circle (2pt);
        \fill[black] (B) circle (2pt);
        \fill[black] (C) circle (2pt);
    \end{tikzpicture}}$ \hfill \\[0.8em]
 $\raisebox{-0.16cm}{\begin{tikzpicture}[scale=1]
        \coordinate (A) at (0,0);
        \coordinate (B) at (1/2,0);
        \coordinate (C) at (1,0);
        \draw[thick, black] (0.5,0) ellipse (0.75cm and 0.25cm);
        \draw[thick, black] (0.75,0) ellipse (0.41cm and 0.19cm);
        \draw[thick] (A) -- (B) -- (C);
        \fill[black] (A) circle (2pt);
        \fill[black] (B) circle (2pt);
        \fill[black] (C) circle (2pt);
    \end{tikzpicture}} \leftrightarrow \raisebox{-0.07cm}{\begin{tikzpicture}[scale=1]
        \coordinate (A) at (0,0);
        \coordinate (B) at (1/2,0);
        \coordinate (C) at (1,0);
        \draw[thick] (A) -- node {\al} (B) -- (C);
        \fill[black] (A) circle (2pt);
        \fill[black] (B) circle (2pt);
        \fill[black] (C) circle (2pt);
    \end{tikzpicture}}$, \hfill & $\raisebox{-0.16cm}{\begin{tikzpicture}[scale=1]
        \coordinate (A) at (0,0);
        \coordinate (B) at (1/2,0);
        \coordinate (C) at (1,0);
        \draw[thick, black] (0.5,0) ellipse (0.75cm and 0.25cm);
        \draw[thick, black] (0.25,0) ellipse (0.41cm and 0.19cm);
        \draw[thick] (A) -- (B) -- (C);
        \fill[black] (A) circle (2pt);
        \fill[black] (B) circle (2pt);
        \fill[black] (C) circle (2pt);
        \draw[black,thick] (A) circle (3pt);
    \end{tikzpicture}} \leftrightarrow \raisebox{-0.07cm}{\begin{tikzpicture}[scale=1]
        \coordinate (A) at (0,0);
        \coordinate (B) at (1/2,0);
        \coordinate (C) at (1,0);
        \draw[thick] (A) -- node {\ar} (B) -- node {\ar} (C);
        \fill[black] (A) circle (2pt);
        \fill[black] (B) circle (2pt);
        \fill[black] (C) circle (2pt);
    \end{tikzpicture}}$\hfill & $\raisebox{-0.16cm}{\begin{tikzpicture}[scale=1]
        \coordinate (A) at (0,0);
        \coordinate (B) at (1/2,0);
        \coordinate (C) at (1,0);
        \draw[thick, black] (0.5,0) ellipse (0.75cm and 0.25cm);
        \draw[thick, black] (0.25,0) ellipse (0.41cm and 0.19cm);
        \draw[thick] (A) -- (B) -- (C);
        \fill[black] (A) circle (2pt);
        \fill[black] (B) circle (2pt);
        \fill[black] (C) circle (2pt);
        \draw[black,thick] (B) circle (3pt);
    \end{tikzpicture}} \leftrightarrow \raisebox{-0.07cm}{\begin{tikzpicture}[scale=1]
        \coordinate (A) at (0,0);
        \coordinate (B) at (1/2,0);
        \coordinate (C) at (1,0);
        \draw[thick] (A) -- node {\al} (B) -- node {\ar} (C);
        \fill[black] (A) circle (2pt);
        \fill[black] (B) circle (2pt);
        \fill[black] (C) circle (2pt);
    \end{tikzpicture}}$ \hfill & $\raisebox{-0.16cm}{\begin{tikzpicture}[scale=1]
        \coordinate (A) at (0,0);
        \coordinate (B) at (1/2,0);
        \coordinate (C) at (1,0);
        \draw[thick, black] (0.5,0) ellipse (0.75cm and 0.25cm);
         \draw[thick, black] (0.75,0) ellipse (0.41cm and 0.19cm);
        \draw[thick] (A) -- (B) -- (C);
        \fill[black] (A) circle (2pt);
        \fill[black] (B) circle (2pt);
        \fill[black] (C) circle (2pt);
        \draw[black,thick] (B) circle (3pt);
    \end{tikzpicture}} \leftrightarrow \raisebox{-0.07cm}{\begin{tikzpicture}[scale=1]
        \coordinate (A) at (0,0);
        \coordinate (B) at (1/2,0);
        \coordinate (C) at (1,0);
        \draw[thick] (A) -- node {\al}  (B) -- node {\ar} (C);
        \fill[black] (A) circle (2pt);
        \fill[black] (B) circle (2pt);
        \fill[black] (C) circle (2pt);
    \end{tikzpicture}}$ \hfill \\[0.8em]
    $\raisebox{-0.16cm}{\begin{tikzpicture}[scale=1]
        \coordinate (A) at (0,0);
        \coordinate (B) at (1/2,0);
        \coordinate (C) at (1,0);
        \draw[thick, black] (0.5,0) ellipse (0.75cm and 0.25cm);
         \draw[thick, black] (0.75,0) ellipse (0.41cm and 0.19cm);
        \draw[thick] (A) -- (B) -- (C);
        \fill[black] (A) circle (2pt);
        \fill[black] (B) circle (2pt);
        \fill[black] (C) circle (2pt);
        \draw[black,thick] (C) circle (3pt);
    \end{tikzpicture}} \leftrightarrow \raisebox{-0.07cm}{\begin{tikzpicture}[scale=1]
        \coordinate (A) at (0,0);
        \coordinate (B) at (1/2,0);
        \coordinate (C) at (1,0);
        \draw[thick] (A) -- node {\al}  (B) -- node {\al} (C);
        \fill[black] (A) circle (2pt);
        \fill[black] (B) circle (2pt);
        \fill[black] (C) circle (2pt);
    \end{tikzpicture}}$, \hfill & $\raisebox{-0.16cm}{\begin{tikzpicture}[scale=1]
        \coordinate (A) at (0,0);
        \coordinate (B) at (1/2,0);
        \coordinate (C) at (1,0);
        \draw[thick, black] (0.5,0) ellipse (0.75cm and 0.25cm);
        \draw[thick] (A) -- (B) -- (C);
        \fill[black] (A) circle (2pt);
        \fill[black] (B) circle (2pt);
        \fill[black] (C) circle (2pt);
        \draw[black,thick] (A) circle (3pt);
        \draw[black,thick] (C) circle (3pt);
    \end{tikzpicture}} \leftrightarrow \raisebox{-0.07cm}{\begin{tikzpicture}[scale=1]
        \coordinate (A) at (0,0);
        \coordinate (B) at (1/2,0);
        \coordinate (C) at (1,0);
        \draw[thick] (A) -- node {\ar}  (B) -- node {\al} (C);
        \fill[black] (A) circle (2pt);
        \fill[black] (B) circle (2pt);
        \fill[black] (C) circle (2pt);
    \end{tikzpicture}}$, \hfill &
    $\raisebox{-0.1cm}{\begin{tikzpicture}[scale=1]
        \coordinate (A) at (0,0);
        \coordinate (B) at (1/2,0);
        \coordinate (C) at (1,0);
        \draw[black,thick] (A) circle (3pt);
        \draw[thick, black] (0.25,0) ellipse (0.41cm and 0.19cm);
        \draw[thick] (A) -- (B) -- (C);
        \draw[black,thick] (C) circle (3pt);
        \fill[black] (A) circle (2pt);
        \fill[black] (B) circle (2pt);
        \fill[black] (C) circle (2pt);
    \end{tikzpicture}} \leftrightarrow \raisebox{-0.07cm}{\begin{tikzpicture}[scale=1]
        \coordinate (A) at (0,0);
        \coordinate (B) at (1/2,0);
        \coordinate (C) at (1,0);
        \draw[thick] (A) -- node {\ar} (B);
        \draw[thick,dotted] (B) -- (C);
        \fill[black] (A) circle (2pt);
        \fill[black] (B) circle (2pt);
        \fill[black] (C) circle (2pt);
    \end{tikzpicture}}$, \hfill & $\raisebox{-0.1cm}{\begin{tikzpicture}[scale=1]
        \coordinate (A) at (0,0);
        \coordinate (B) at (1/2,0);
        \coordinate (C) at (1,0);
        \draw[black,thick] (B) circle (3pt);
        \draw[thick, black] (0.25,0) ellipse (0.41cm and 0.19cm);
        \draw[thick] (A) -- (B) -- (C);
        \draw[black,thick] (C) circle (3pt);
        \fill[black] (A) circle (2pt);
        \fill[black] (B) circle (2pt);
        \fill[black] (C) circle (2pt);
    \end{tikzpicture}} \leftrightarrow \raisebox{-0.07cm}{\begin{tikzpicture}[scale=1]
        \coordinate (A) at (0,0);
        \coordinate (B) at (1/2,0);
        \coordinate (C) at (1,0);
        \draw[thick] (A) -- node {\al} (B);
        \draw[thick,dotted] (B) -- (C);
        \fill[black] (A) circle (2pt);
        \fill[black] (B) circle (2pt);
        \fill[black] (C) circle (2pt);
    \end{tikzpicture}}$,\hfill  \\[0.8em]
      $\raisebox{-0.1cm}{\begin{tikzpicture}[scale=1]
        \coordinate (A) at (0,0);
        \coordinate (B) at (1/2,0);
        \coordinate (C) at (1,0);
        \draw[thick, black] (0.75,0) ellipse (0.41cm and 0.19cm);
        \draw[thick] (A) -- (B) -- (C);
        \draw[black,thick] (A) circle (3pt);
        \draw[black,thick] (B) circle (3pt);
        \fill[black] (A) circle (2pt);
        \fill[black] (B) circle (2pt);
        \fill[black] (C) circle (2pt);
    \end{tikzpicture}} \leftrightarrow \raisebox{-0.07cm}{\begin{tikzpicture}[scale=1]
        \coordinate (A) at (0,0);
        \coordinate (B) at (1/2,0);
        \coordinate (C) at (1,0);
        \draw[thick,dotted] (A) -- (B);
        \draw[thick] (B) -- node {\ar} (C);
        \fill[black] (A) circle (2pt);
        \fill[black] (B) circle (2pt);
        \fill[black] (C) circle (2pt);
    \end{tikzpicture}}$, \hfill & $\raisebox{-0.1cm}{\begin{tikzpicture}[scale=1]
        \coordinate (A) at (0,0);
        \coordinate (B) at (1/2,0);
        \coordinate (C) at (1,0);
        \draw[thick, black] (0.75,0) ellipse (0.41cm and 0.19cm);
        \draw[thick] (A) -- (B) -- (C);
        \draw[black,thick] (A) circle (3pt);
        \draw[black,thick] (C) circle (3pt);
        \fill[black] (A) circle (2pt);
        \fill[black] (B) circle (2pt);
        \fill[black] (C) circle (2pt);
    \end{tikzpicture}} \leftrightarrow \raisebox{-0.07cm}{\begin{tikzpicture}[scale=1]
        \coordinate (A) at (0,0);
        \coordinate (B) at (1/2,0);
        \coordinate (C) at (1,0);
        \draw[thick,dotted] (A) -- (B);
        \draw[thick] (B) -- node {\al} (C);
        \fill[black] (A) circle (2pt);
        \fill[black] (B) circle (2pt);
        \fill[black] (C) circle (2pt);
    \end{tikzpicture}}$. \hfill &  \hfill& \hfill
\end{tabular}
\end{center}
The two-cycle has the following cut tubings and corresponding decorated orientations
\begin{center}
\begin{tabular}{cccc}
$\raisebox{-0.29cm}{\begin{tikzpicture}[scale=0.8]
\fill[black] (0,-1) circle (2pt);
\fill[black] (1,-1) circle (2pt);
\draw[thick] (0,-1) to[out=90,in=90] (1,-1);
\draw[thick] (0,-1) to[out=-90,in=180+90] (1,-1);
\draw[thick, black] (0.5,-1) ellipse (0.8cm and 0.5cm);
\end{tikzpicture}} \leftrightarrow \raisebox{-0.2cm}{\begin{tikzpicture}[scale=0.8]
\fill[black] (0,-1) circle (2pt);
\fill[black] (1,-1) circle (2pt);
\draw[thick] (0,-1) to[out=90,in=90] (1,-1);
\draw[thick] (0,-1) to[out=-90,in=180+90] (1,-1);
\end{tikzpicture}}$, \hfill & $\raisebox{-0.2cm}{\begin{tikzpicture}[scale=0.8]
\fill[black] (0,-1) circle (2pt);
\fill[black] (1,-1) circle (2pt);
\draw[thick] (0,-1) to[out=90,in=90] (1,-1);
\draw[thick] (0,-1) to[out=-90,in=180+90] (1,-1);
\draw[thick] (0-0.2,-1) to[out=90,in=90] (1+0.2,-1);
\draw[thick] (0+0.2,-1) to[out=90,in=90] (1-0.2,-1);
\draw[thick] (1-0.2,-1) to[out=90+180,in=180] (1,-1.2) to[out=0,in=90+180] (1+0.2,-1);
\draw[thick] (0-0.2,-1) to[out=90+180,in=180] (0,-1.2) to[out=0,in=90+180] (0+0.2,-1);
\end{tikzpicture}} \leftrightarrow \raisebox{-0.2cm}{\begin{tikzpicture}[scale=0.8]
\fill[black] (0,-1) circle (2pt);
\fill[black] (1,-1) circle (2pt);
\draw[thick] (0,-1) to[out=90,in=90] (1,-1);
\draw[thick,dotted] (0,-1) to[out=-90,in=180+90] (1,-1);
\end{tikzpicture}}$, \hfill & $\raisebox{-0.3cm}{\begin{tikzpicture}[scale=0.8]
\fill[black] (0,-1) circle (2pt);
\fill[black] (1,-1) circle (2pt);
\draw[thick] (0,-1) to[out=90,in=90] (1,-1);
\draw[thick] (0,-1) to[out=-90,in=180+90] (1,-1);
\draw[thick] (0-0.2,-1) to[out=-90,in=180+90] (1+0.2,-1);
\draw[thick] (0+0.2,-1) to[out=-90,in=180+90] (1-0.2,-1);
\draw[thick] (1-0.2,-1) to[out=90,in=180] (1,-0.8) to[out=0,in=90] (1+0.2,-1);
\draw[thick] (0-0.2,-1) to[out=90,in=180] (0,-0.8) to[out=0,in=90] (0+0.2,-1);
\end{tikzpicture}} \leftrightarrow \raisebox{-0.2cm}{\begin{tikzpicture}[scale=0.8]
\fill[black] (0,-1) circle (2pt);
\fill[black] (1,-1) circle (2pt);
\draw[thick,dotted] (0,-1) to[out=90,in=90] (1,-1);
\draw[thick] (0,-1) to[out=-90,in=180+90] (1,-1);
\end{tikzpicture}}$, \hfill & $\raisebox{-0.2cm}{\begin{tikzpicture}[scale=0.8]
\fill[black] (0,-1) circle (2pt);
\fill[black] (1,-1) circle (2pt);
\draw[black,thick] (0,-1) circle (4pt);
\draw[black,thick] (1,-1) circle (4pt);
\draw[thick] (0,-1) to[out=90,in=90] (1,-1);
\draw[thick] (0,-1) to[out=-90,in=180+90] (1,-1);
\end{tikzpicture}} \leftrightarrow \raisebox{-0.2cm}{\begin{tikzpicture}[scale=0.8]
\fill[black] (0,-1) circle (2pt);
\fill[black] (1,-1) circle (2pt);
\draw[thick,dotted] (0,-1) to[out=90,in=90] (1,-1);
\draw[thick,dotted] (0,-1) to[out=-90,in=180+90] (1,-1);
\end{tikzpicture}}$, \hfill  \\[0.8em]
$\raisebox{-0.29cm}{\begin{tikzpicture}[scale=0.8]
\fill[black] (0,-1) circle (2pt);
\fill[black] (1,-1) circle (2pt);
\draw[thick] (0,-1) to[out=90,in=90] (1,-1);
\draw[thick] (0,-1) to[out=-90,in=180+90] (1,-1);
\draw[thick, black] (0.5,-1) ellipse (0.8cm and 0.5cm);
\draw[black,thick] (0,-1) circle (4pt);
\end{tikzpicture}} \leftrightarrow \raisebox{-0.3cm}{\begin{tikzpicture}[scale=0.8]
\fill[black] (0,-1) circle (2pt);
\fill[black] (1,-1) circle (2pt);
\draw[thick] (0,-1) to[out=90,in=90] node {\ar} (1,-1);
\draw[thick] (0,-1) to[out=-90,in=180+90] node {\ar} (1,-1);
\end{tikzpicture}}$, \hfill & $\raisebox{-0.29cm}{\begin{tikzpicture}[scale=0.8]
\fill[black] (0,-1) circle (2pt);
\fill[black] (1,-1) circle (2pt);
\draw[thick] (0,-1) to[out=90,in=90] (1,-1);
\draw[thick] (0,-1) to[out=-90,in=180+90] (1,-1);
\draw[thick, black] (0.5,-1) ellipse (0.8cm and 0.5cm);
\draw[black,thick] (1,-1) circle (4pt);
\end{tikzpicture}} \leftrightarrow \raisebox{-0.3cm}{\begin{tikzpicture}[scale=0.8]
\fill[black] (0,-1) circle (2pt);
\fill[black] (1,-1) circle (2pt);
\draw[thick] (0,-1) to[out=90,in=90] node {\al} (1,-1);
\draw[thick] (0,-1) to[out=-90,in=180+90] node {\al} (1,-1);
\end{tikzpicture}}$, \hfill & $\raisebox{-0.3cm}{\begin{tikzpicture}[scale=0.8]
\fill[black] (0,-1) circle (2pt);
\fill[black] (1,-1) circle (2pt);
\draw[thick] (0,-1) to[out=90,in=90] (1,-1);
\draw[thick] (0,-1) to[out=-90,in=180+90] (1,-1);
\draw[thick] (0-0.2,-1) to[out=-90,in=180+90] (1+0.2,-1);
\draw[thick] (0+0.2,-1) to[out=-90,in=180+90] (1-0.2,-1);
\draw[thick] (1-0.2,-1) to[out=90,in=180] (1,-0.8) to[out=0,in=90] (1+0.2,-1);
\draw[thick] (0-0.2,-1) to[out=90,in=180] (0,-0.8) to[out=0,in=90] (0+0.2,-1);
\draw[black,thick] (0,-1) circle (4pt);
\end{tikzpicture}} \leftrightarrow \raisebox{-0.27cm}{\begin{tikzpicture}[scale=0.8]
\fill[black] (0,-1) circle (2pt);
\fill[black] (1,-1) circle (2pt);
\draw[thick,dotted] (0,-1) to[out=90,in=90] (1,-1);
\draw[thick] (0,-1) to[out=-90,in=180+90] node {\ar} (1,-1);
\end{tikzpicture}}$, \hfill & $\raisebox{-0.3cm}{\begin{tikzpicture}[scale=0.8]
\fill[black] (0,-1) circle (2pt);
\fill[black] (1,-1) circle (2pt);
\draw[thick] (0,-1) to[out=90,in=90] (1,-1);
\draw[thick] (0,-1) to[out=-90,in=180+90] (1,-1);
\draw[thick] (0-0.2,-1) to[out=-90,in=180+90] (1+0.2,-1);
\draw[thick] (0+0.2,-1) to[out=-90,in=180+90] (1-0.2,-1);
\draw[thick] (1-0.2,-1) to[out=90,in=180] (1,-0.8) to[out=0,in=90] (1+0.2,-1);
\draw[thick] (0-0.2,-1) to[out=90,in=180] (0,-0.8) to[out=0,in=90] (0+0.2,-1);
\draw[black,thick] (1,-1) circle (4pt);
\end{tikzpicture}} \leftrightarrow \raisebox{-0.27cm}{\begin{tikzpicture}[scale=0.8]
\fill[black] (0,-1) circle (2pt);
\fill[black] (1,-1) circle (2pt);
\draw[thick,dotted] (0,-1) to[out=90,in=90] (1,-1);
\draw[thick] (0,-1) to[out=-90,in=180+90] node {\al} (1,-1);
\end{tikzpicture}}$, \hfill  \\[0.8em]
$\raisebox{-0.2cm}{\begin{tikzpicture}[scale=0.8]
\fill[black] (0,-1) circle (2pt);
\fill[black] (1,-1) circle (2pt);
\draw[thick] (0,-1) to[out=90,in=90] (1,-1);
\draw[thick] (0,-1) to[out=-90,in=180+90] (1,-1);
\draw[thick] (0-0.2,-1) to[out=90,in=90] (1+0.2,-1);
\draw[thick] (0+0.2,-1) to[out=90,in=90] (1-0.2,-1);
\draw[thick] (1-0.2,-1) to[out=90+180,in=180] (1,-1.2) to[out=0,in=90+180] (1+0.2,-1);
\draw[thick] (0-0.2,-1) to[out=90+180,in=180] (0,-1.2) to[out=0,in=90+180] (0+0.2,-1);
\draw[black,thick] (0,-1) circle (4pt);
\end{tikzpicture}} \leftrightarrow \raisebox{-0.2cm}{\begin{tikzpicture}[scale=0.8]
\fill[black] (0,-1) circle (2pt);
\fill[black] (1,-1) circle (2pt);
\draw[thick] (0,-1) to[out=90,in=90] node {\ar} (1,-1);
\draw[thick,dotted] (0,-1) to[out=-90,in=180+90] (1,-1);
\end{tikzpicture}}$, \hfill & $\raisebox{-0.2cm}{\begin{tikzpicture}[scale=0.8]
\fill[black] (0,-1) circle (2pt);
\fill[black] (1,-1) circle (2pt);
\draw[thick] (0,-1) to[out=90,in=90] (1,-1);
\draw[thick] (0,-1) to[out=-90,in=180+90] (1,-1);
\draw[thick] (0-0.2,-1) to[out=90,in=90] (1+0.2,-1);
\draw[thick] (0+0.2,-1) to[out=90,in=90] (1-0.2,-1);
\draw[thick] (1-0.2,-1) to[out=90+180,in=180] (1,-1.2) to[out=0,in=90+180] (1+0.2,-1);
\draw[thick] (0-0.2,-1) to[out=90+180,in=180] (0,-1.2) to[out=0,in=90+180] (0+0.2,-1);
\draw[black,thick] (1,-1) circle (4pt);
\end{tikzpicture}} \leftrightarrow \raisebox{-0.2cm}{\begin{tikzpicture}[scale=0.8]
\fill[black] (0,-1) circle (2pt);
\fill[black] (1,-1) circle (2pt);
\draw[thick] (0,-1) to[out=90,in=90] node {\al} (1,-1);
\draw[thick,dotted] (0,-1) to[out=-90,in=180+90] (1,-1);
\end{tikzpicture}}$. \hfill &  \hfill &  \hfill  \\[0.8em]
\end{tabular}
\end{center}
As can be seen by the above examples multiple cut tubings can lead to the same decorated orientation. Naively, a graph can be decorated in $4^{|E_G|}$ many ways, however, as demonstrated by the two-cycle, not all graph decorations arise from cut-tubings, which in this case produce $10$ decorated orientations as opposed to the naive counting $16=4 \times 4$. In the next section we show how the decorated orientations can be defined without reference to cut tubings. As we will see, the subset of decorated orientations arising from the cut tubings is then selected by a simple {\it acyclic} rule.

\subsection{Counting the kinematic flow}
We define the set of {\it decorated orientations} of $G$ to be the set of all graphs obtained by assigning to each edge $e \in E_G$ one of the following decorations: $\begin{tikzpicture}[scale=1]
        \coordinate (A) at (0,0);
        \coordinate (B) at (1/2,0);
        \coordinate (C) at (1,0);
        \coordinate (D) at (3/2,0);
        \draw[thick] (A) -- (B);
        \fill[black] (A) circle (2pt);
        \fill[black] (B) circle (2pt);
    \end{tikzpicture}$, $\begin{tikzpicture}[scale=1]
        \coordinate (A) at (0,0);
        \coordinate (B) at (1/2,0);
        \coordinate (C) at (1,0);
        \coordinate (D) at (3/2,0);
        \draw[thick,dotted] (A) -- (B);
        \fill[black] (A) circle (2pt);
        \fill[black] (B) circle (2pt);
    \end{tikzpicture}$, $\raisebox{-0.11cm}{\begin{tikzpicture}[scale=1]
        \coordinate (A) at (0,0);
        \coordinate (B) at (1/2,0);
        \coordinate (C) at (1,0);
        \coordinate (D) at (3/2,0);
        \draw[thick] (A) --  node {\al} (B);
        \fill[black] (A) circle (2pt);
        \fill[black] (B) circle (2pt);
    \end{tikzpicture}}$ or $\raisebox{-0.11cm}{\begin{tikzpicture}[scale=1]
        \coordinate (A) at (0,0);
        \coordinate (B) at (1/2,0);
        \coordinate (C) at (1,0);
        \coordinate (D) at (3/2,0);
        \draw[thick] (A) --  node {\ar} (B);
        \fill[black] (A) circle (2pt);
        \fill[black] (B) circle (2pt);
    \end{tikzpicture}}$. The first two we refer to as solid and broken edges respectively whereas the last two we refer to as oriented edges. We denote the set of all decorated orientations of the graph by $\text{Dec}(G)$. It is clear we have $|\text{Dec}(G)|=4^{|E_G|}$.
    
\begin{figure}[]
\centering
\includegraphics[scale=0.9]{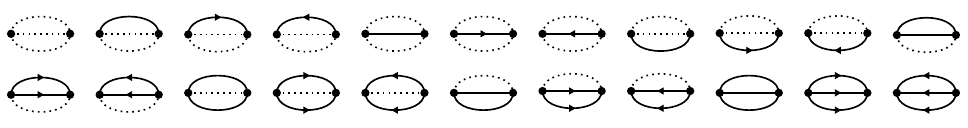}
\caption{The $22$ acyclic decorated orientations of the two-loop sunrise.}
\label{fig:1}
\end{figure}

\begin{figure}[]
\centering
\includegraphics[scale=0.95]{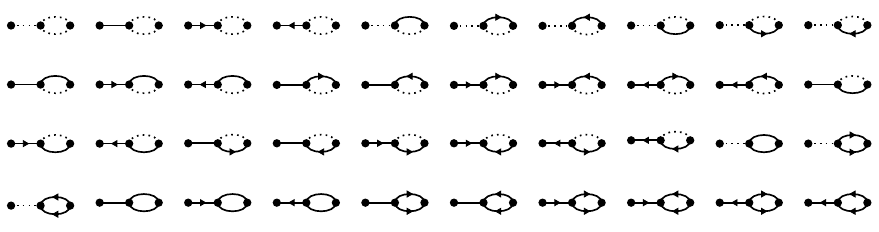}
\caption{The $40$ acyclic decorated orientations of the one-loop frying pan.}
\label{fig:5}
\end{figure}

\begin{figure}[]
\centering
\includegraphics[scale=1]{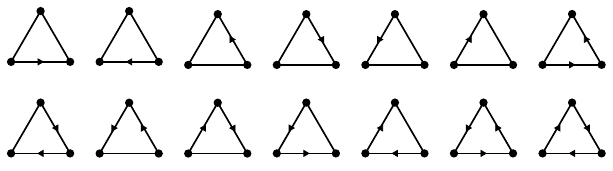}
\caption{The $14$ decorated orientations of the three cycle that fail to be acyclic.}
\label{fig:bad1}
\end{figure}

To make connection to the set of decorated orientations which arise from considering the cut tubings of the last section we must introduce a notion of acyclicity of a decorated orientation. A decorated orientation of a graph $G$ is {\it acyclic} if the oriented graph obtained by deleting all broken edges and contracting all solid edges is acyclic in the usual sense of an oriented graph. Where an oriented graph is said to be acyclic if it contains no cycles with all edges oriented in the same direction. Examples of acyclic decorated orientations for various graphs are displayed in Fig.~\ref{fig:1} through to Fig.~\ref{fig:5}. Examples of decorated orientations which fail to be acyclic are given in Fig.~\ref{fig:bad1}. We denote the set of all {\it acyclic decorated orientations} by $\text{aDec}(G)$. It is immediate from the definition that the cardinality of the set $\text{aDec}(G)$ is given by
\begin{align}
|\text{aDec}(G)| = \sum_{{\bf e} \subset E_G}  \sum_{{\bf e'} \subset {\bf e}} |\text{aDir}(G_{{\bf e},{\bf e'}})|.
\end{align}
Where the graph $G_{{\bf e}, {\bf e'}}$ is obtained from the graph $G$ by deleting all edges in the set $E_G \setminus {\bf e}$ and contracting all edges in the set ${\bf e'}$. The factor $|\text{aDir}(G_{{\bf e},{\bf e'}})|$ counts the number of acyclic orientations of the graph $G_{{\bf e}, {\bf e'}}$. This is a well known graph invariant given by the Tutte polynomial $T(G_{{\bf e},{\bf e'}};{x,y})$ evaluated at $x=2$ and $y=0$, that is
\begin{align}
|\text{aDir}(G_{{\bf e},{\bf e'}})| = T(G_{{\bf e},{\bf e'}};2,0).
\end{align}
Some examples for the number of acyclic decorated orientations for various graphs are given in Table \ref{tab:1}. 
\begin{table}[h!]
\centering
\begin{tabular}{|c|c|c|c|c|c|}
\hline
$n=$ & 1 & 2 & 3 & 4 & 5\\ \hline
$|\text{aDir}(P_n)|$ & 1 & 4 & 16 & 64 & 256 \\ \hline
$|\text{aDir}(C_n)|$ & 2 & 10 & 50 & 226 & 962 \\ \hline
$|\text{aDir}(W_n)|$ & 1 & 8 & 118 & 1688 & 22030 \\ \hline
$|\text{aDir}(K_n)|$ & 1 & 4 & 50 & 1688 & 142624 \\ \hline
\end{tabular}
\caption{The number of acyclic decorated orientations for the path graph $P_n$, cycle graph $C_n$, wheel graph $W_n$ and complete graph $K_n$ on $n$ vertices.}
\label{tab:1}
\end{table}

Remarkably, we find that the cardinality of the set of acyclic decorated orientations for a graph $G$ is equal to the number of basis functions appearing in the kinematic flow discovered in \cite{Arkani-Hamed:2023kig,Arkani-Hamed:2023bsv}. This leads us to conjecture the following 
\begin{center}
\boxed{\text{ {\it The functions appearing in the kinematic flow for the graph $G$ are counted by $|\text{aDec}(G)|$. }}}
\end{center}
Our rule for determining labels for the set of functions appearing in the kinematic flow for an arbitrary graph $G$ can be mapped to that of \cite{Baumann:2024mvm}. The kinematic flow and its connection to the combinatorics of graph tubings has also been studied in \cite{Grimm:2024mbw,Grimm:2025zhv,He:2024olr,Fan:2024iek,Hang:2024xas}. We leave a more detailed investigation of the connection between decorated orientations and the kinematic flow to future work.

\begin{figure}[]
\centering
\includegraphics[scale=0.9]{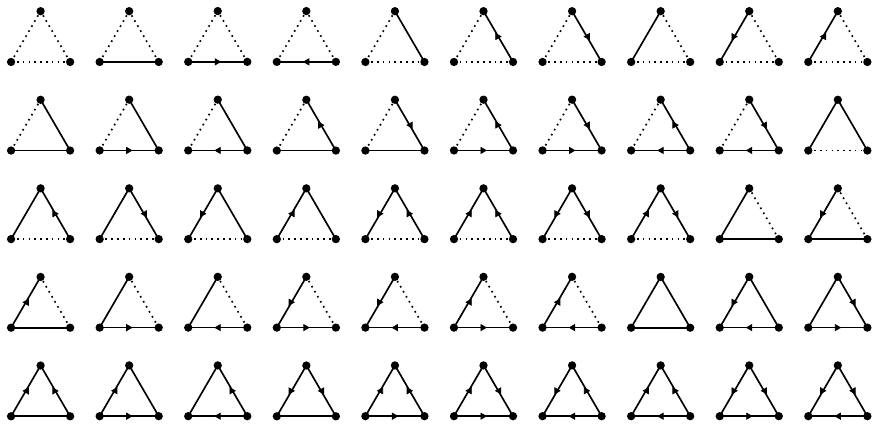}
\caption{The $50$ acyclic decorated orientations of the three cycle.}
\end{figure}

%========================================================================================
%========================================================================================
\section{Conclusion}
\label{sec:conc}
In this paper we have explored the similarities between the edge and vertex centric notions of tubings which have appeared in the physics and mathematics literature referred to here as binary and unary tubes respectively. The binary tubes are the building blocks for computing the wavefunction of the universe, whereas the unary tubes are the building blocks of amplitubes. Although taking different forms we have shown that the wavefunction coefficients and amplitubes are intimately connected. In \eqref{eq:edge_expand} we presented a formula for the wavefunction coefficient of a graph $G$ as a sum over $2^{|E_G|}$ many terms, each associated to cutting a certain subset of edges of the graph. Each term appearing in the sum is then given by a product of amplitubes, one for each connected component of the graph with the specified edges cut.  Furthermore, we have shown how the amplitubes can natrually be expanded as a sum over orientations of the underlying graph. Combining this observation with \eqref{eq:edge_expand} resulted in a general formula for the wavefunction coefficient as a sum over {decorated amplitubes} with edges decorated by either of the three options: $\begin{tikzpicture}[scale=1]
        \coordinate (A) at (0,0);
        \coordinate (B) at (1/2,0);
        \coordinate (C) at (1,0);
        \coordinate (D) at (3/2,0);
        \draw[thick,dotted] (A) -- (B);
        \fill[black] (A) circle (2pt);
        \fill[black] (B) circle (2pt);
    \end{tikzpicture}$, $\raisebox{-0.11cm}{\begin{tikzpicture}[scale=1]
        \coordinate (A) at (0,0);
        \coordinate (B) at (1/2,0);
        \coordinate (C) at (1,0);
        \coordinate (D) at (3/2,0);
        \draw[thick] (A) --  node {\al} (B);
        \fill[black] (A) circle (2pt);
        \fill[black] (B) circle (2pt);
    \end{tikzpicture}}$ or $\raisebox{-0.11cm}{\begin{tikzpicture}[scale=1]
        \coordinate (A) at (0,0);
        \coordinate (B) at (1/2,0);
        \coordinate (C) at (1,0);
        \coordinate (D) at (3/2,0);
        \draw[thick] (A) --  node {\ar} (B);
        \fill[black] (A) circle (2pt);
        \fill[black] (B) circle (2pt);
    \end{tikzpicture}}$. The expansion we provide for the wavefunction in terms of decorated amplitubes, our formula \eqref{eq:main}, matches previous results in the literature where, in \cite{Arkani-Hamed:2017fdk} it is referred to as the {\it bulk time integral} representation and in \cite{Fevola:2024nzj} as the partial fractioning formula.
    
Our results \eqref{eq:edge_expand} and \eqref{eq:main} for the wavefunction coefficient suggested a new hybrid definition of tubing which makes use of both binary and unary tubes together which we refer to as cut tubings, see also \cite{De:2024zic}. Given a cut tubing we showed how to assign a {\it decorated orientation} to the underlying graph, where edges recieve one of the following four decorations: $\begin{tikzpicture}[scale=1]
        \coordinate (A) at (0,0);
        \coordinate (B) at (1/2,0);
        \coordinate (C) at (1,0);
        \coordinate (D) at (3/2,0);
        \draw[thick] (A) -- (B);
        \fill[black] (A) circle (2pt);
        \fill[black] (B) circle (2pt);
    \end{tikzpicture}$, $\begin{tikzpicture}[scale=1]
        \coordinate (A) at (0,0);
        \coordinate (B) at (1/2,0);
        \coordinate (C) at (1,0);
        \coordinate (D) at (3/2,0);
        \draw[thick,dotted] (A) -- (B);
        \fill[black] (A) circle (2pt);
        \fill[black] (B) circle (2pt);
    \end{tikzpicture}$, $\raisebox{-0.11cm}{\begin{tikzpicture}[scale=1]
        \coordinate (A) at (0,0);
        \coordinate (B) at (1/2,0);
        \coordinate (C) at (1,0);
        \coordinate (D) at (3/2,0);
        \draw[thick] (A) --  node {\al} (B);
        \fill[black] (A) circle (2pt);
        \fill[black] (B) circle (2pt);
    \end{tikzpicture}}$ or $\raisebox{-0.11cm}{\begin{tikzpicture}[scale=1]
        \coordinate (A) at (0,0);
        \coordinate (B) at (1/2,0);
        \coordinate (C) at (1,0);
        \coordinate (D) at (3/2,0);
        \draw[thick] (A) --  node {\ar} (B);
        \fill[black] (A) circle (2pt);
        \fill[black] (B) circle (2pt);
    \end{tikzpicture}}$. Remarkably, we found that the set of {\it acyclic} decorated orientations counts the number of basis functions in the kinematic flow \cite{Arkani-Hamed:2023kig,Arkani-Hamed:2023bsv,Baumann:2024mvm} matching previous results of \cite{Baumann:2024mvm}. We leave a more systematic study of the connection between acyclic decorated orientations and the kinematic flow to future work.
    
\section*{Acknowledgments}
I would like to thank Tomasz \L ukowski for collaboration on related topics. Also Shannon Glew for preliminary discussions on the paper. 
%%%%%%%%%%%%%%%%%%%%%%%%%%%%%%%%%%%%%%%%%%%
\bibliographystyle{nb}

\bibliography{amplitubes}

\begin{thebibliography}{10}
\ifx\href\asklfhas\newcommand{\href}[2]{#2}\fi
\ifx\arxivref\asklfhas\newcommand{\arxivref}[2]{\href{http://arxiv.org/abs/#1}{#2}}\fi
\ifx\doiref\asklfhas\newcommand{\doiref}[2]{\href{http://dx.doi.org/#1}{#2}}\fi
\raggedright
\small
\parskip 0pt

\bibitem{Arkani-Hamed:2017fdk}
N.~Arkani-Hamed, P.~Benincasa and A.~Postnikov,
\textit{``{Cosmological Polytopes and the Wavefunction of the Universe}''},
\texttt{\arxivref{1709.02813}{arxiv:1709.02813}}.

\bibitem{carr2006coxeter}
M.~Carr and S.~L.~Devadoss,
\textit{``Coxeter complexes and graph-associahedra''},
\textsf{Topology~and~its~Applications~153,~2155~(2006)}.

\bibitem{Glew:2025otn}
R.~Glew and T.~Lukowski,
\textit{``{Amplitubes: Graph Cosmohedra}''},
\texttt{\arxivref{2502.17564}{arxiv:2502.17564}}.

\bibitem{Arkani-Hamed:2017tmz}
N.~Arkani-Hamed, Y.~Bai and T.~Lam,
\textit{``{Positive Geometries and Canonical Forms}''},
\textsf{\doiref{10.1007/JHEP11(2017)039}{JHEP~1711,~039~(2017)}},
\texttt{\arxivref{1703.04541}{arxiv:1703.04541}}.

\bibitem{Benincasa:2024leu}
P.~Benincasa and G.~Dian,
\textit{``{The Geometry of Cosmological Correlators}''},
\texttt{\arxivref{2401.05207}{arxiv:2401.05207}}.

\bibitem{De:2023xue}
S.~De and A.~Pokraka,
\textit{``{Cosmology meets cohomology}''},
\textsf{\doiref{10.1007/JHEP03(2024)156}{JHEP~2403,~156~(2024)}},
\texttt{\arxivref{2308.03753}{arxiv:2308.03753}}.

\bibitem{Benincasa:2024lxe}
P.~Benincasa and F.~Vaz\~ao,
\textit{``{The Asymptotic Structure of Cosmological Integrals}''},
\texttt{\arxivref{2402.06558}{arxiv:2402.06558}}.

\bibitem{De:2024zic}
S.~De and A.~Pokraka,
\textit{``{A physical basis for cosmological correlators from cuts}''},
\texttt{\arxivref{2411.09695}{arxiv:2411.09695}}.

\bibitem{Arkani-Hamed:2023bsv}
N.~Arkani-Hamed, D.~Baumann, A.~Hillman, A.~Joyce, H.~Lee and G.~L.~Pimentel,
\textit{``{Kinematic Flow and the Emergence of Time}''},
\texttt{\arxivref{2312.05300}{arxiv:2312.05300}}.

\bibitem{Arkani-Hamed:2023kig}
N.~Arkani-Hamed, D.~Baumann, A.~Hillman, A.~Joyce, H.~Lee and G.~L.~Pimentel,
\textit{``{Differential Equations for Cosmological Correlators}''},
\texttt{\arxivref{2312.05303}{arxiv:2312.05303}}.

\bibitem{Benincasa:2020aoj}
P.~Benincasa, A.~J.~McLeod and C.~Vergu,
\textit{``{Steinmann Relations and the Wavefunction of the Universe}''},
\textsf{\doiref{10.1103/PhysRevD.102.125004}{Phys.~Rev.~D~102,~125004~(2020)}},
\texttt{\arxivref{2009.03047}{arxiv:2009.03047}}.

\bibitem{Arkani-Hamed:2017mur}
N.~Arkani-Hamed, Y.~Bai, S.~He and G.~Yan,
\textit{``{Scattering Forms and the Positive Geometry of Kinematics, Color and
  the Worldsheet}''},
\textsf{\doiref{10.1007/JHEP05(2018)096}{JHEP~1805,~096~(2018)}},
\texttt{\arxivref{1711.09102}{arxiv:1711.09102}}.

\bibitem{Arkani-Hamed:2023lbd}
N.~Arkani-Hamed, H.~Frost, G.~Salvatori, P.-G.~Plamondon and H.~Thomas,
\textit{``{All Loop Scattering As A Counting Problem}''},
\texttt{\arxivref{2309.15913}{arxiv:2309.15913}}.

\bibitem{He:2020onr}
S.~He, Z.~Li, P.~Raman and C.~Zhang,
\textit{``{Stringy canonical forms and binary geometries from associahedra,
  cyclohedra and generalized permutohedra}''},
\textsf{\doiref{10.1007/JHEP10(2020)054}{JHEP~2010,~054~(2020)}},
\texttt{\arxivref{2005.07395}{arxiv:2005.07395}}.

\bibitem{Arkani-Hamed:2024jbp}
N.~Arkani-Hamed, C.~Figueiredo and F.~Vaz\~ao,
\textit{``{Cosmohedra}''},
\texttt{\arxivref{2412.19881}{arxiv:2412.19881}}.

\bibitem{Melville:2021lst}
S.~Melville and E.~Pajer,
\textit{``{Cosmological Cutting Rules}''},
\textsf{\doiref{10.1007/JHEP05(2021)249}{JHEP~2105,~249~(2021)}},
\texttt{\arxivref{2103.09832}{arxiv:2103.09832}}.

\bibitem{AguiSalcedo:2023nds}
S.~Agui~Salcedo and S.~Melville,
\textit{``{The cosmological tree theorem}''},
\textsf{\doiref{10.1007/JHEP12(2023)076}{JHEP~2312,~076~(2023)}},
\texttt{\arxivref{2308.00680}{arxiv:2308.00680}}.

\bibitem{Fevola:2024nzj}
C.~Fevola, G.~L.~Pimentel, A.-L.~Sattelberger and T.~Westerdijk,
\textit{``{Algebraic Approaches to Cosmological Integrals}''},
\texttt{\arxivref{2410.14757}{arxiv:2410.14757}}.

\bibitem{Baumann:2024mvm}
D.~Baumann, H.~Goodhew and H.~Lee,
\textit{``{Kinematic Flow for Cosmological Loop Integrands}''},
\texttt{\arxivref{2410.17994}{arxiv:2410.17994}}.

\bibitem{Grimm:2024mbw}
T.~W.~Grimm, A.~Hoefnagels and M.~van~Vliet,
\textit{``{Structure and complexity of cosmological correlators}''},
\textsf{\doiref{10.1103/PhysRevD.110.123531}{Phys.~Rev.~D~110,~123531~(2024)}},
\texttt{\arxivref{2404.03716}{arxiv:2404.03716}}.

\bibitem{Grimm:2025zhv}
T.~W.~Grimm, A.~Hoefnagels and M.~van~Vliet,
\textit{``{A Reduction Algorithm for Cosmological Correlators: Cuts,
  Contractions, and Complexity}''},
\texttt{\arxivref{2503.05866}{arxiv:2503.05866}}.

\bibitem{He:2024olr}
S.~He, X.~Jiang, J.~Liu, Q.~Yang and Y.-Q.~Zhang,
\textit{``{Differential equations and recursive solutions for cosmological
  amplitudes}''},
\textsf{\doiref{10.1007/JHEP01(2025)001}{JHEP~2501,~001~(2025)}},
\texttt{\arxivref{2407.17715}{arxiv:2407.17715}}.

\bibitem{Fan:2024iek}
B.~Fan and Z.-Z.~Xianyu,
\textit{``{Cosmological amplitudes in power-law FRW universe}''},
\textsf{\doiref{10.1007/JHEP12(2024)042}{JHEP~2412,~042~(2024)}},
\texttt{\arxivref{2403.07050}{arxiv:2403.07050}}.

\bibitem{Hang:2024xas}
Y.~Hang and C.~Shen,
\textit{``{A Note on Kinematic Flow and Differential Equations for Two-Site
  One-Loop Graph in FRW Spacetime}''},
\texttt{\arxivref{2410.17192}{arxiv:2410.17192}}.

\bibitem{Balduf:2023tls}
P.-H.~Balduf, A.~Cantwell, K.~Ebrahimi-Fard, L.~Nabergall, N.~Olson-Harris and
  K.~Yeats,
\textit{``{Tubings, chord diagrams, and Dyson\textendash{}Schwinger
  equations}''},
\textsf{\doiref{10.1112/jlms.70006}{J.~Lond.~Math.~Soc.~110,~e70006~(2024)}},
\texttt{\arxivref{2302.02019}{arxiv:2302.02019}}.

\bibitem{carr2011pseudograph}
M.~Carr, S.~L.~Devadoss and S.~Forcey,
\textit{``Pseudograph associahedra''},
\textsf{Journal~of~Combinatorial~Theory,~Series~A~118,~2035~(2011)}.

\end{thebibliography}

\end{document}